%% file: lamb_lions.tex
\pdfoutput=1
\documentclass[aps,prl,twocolumn,superscriptaddress,floatfix,nofootinbib]{revtex4-2}

\usepackage{amsmath,amssymb}
\usepackage{graphicx}
\usepackage{bm}
\usepackage{mathtools}
\usepackage{physics}
\usepackage[colorlinks=true,allcolors=blue]{hyperref}

\newcommand{\avg}[1]{\left\langle #1 \right\rangle}

\begin{document}

\title{Capture of a diffusing lamb by three diffusing lions:\\
hidden symmetry and an exact solution}

\author{M.~Biroli}
\affiliation{James Franck Institute, The University of Chicago, Chicago, Illinois 60637, USA}

\date{\today}

\begin{abstract}
    The lamb and lions problem is a classical open problem from stochastic processes with potential applications percolating to any capture process. The statement of the problem is surprisingly simple. A lamb starting at $\ell > 0$, represented by a one-dimensional diffusing random walker, is chased by a pride of $N$ lions starting at the origin, each lion is another independent one-dimensional diffusing random walker. If any lion touches the lamb the process terminates. The question is what is the late-time behavior of the survival probability $S(t)$ of the lamb? The answer is known for $N = 1, 2$ and the leading order is known for $N \gg 1$, but a generic solution for arbitrary $N \geq 3$ has eluded us for the past three decades. In this Letter, we show that the $N = 3$ problem has a hidden symmetry which allows us to provide a closed form solution for $N = 3$. We validate our results with simulations. 
\end{abstract}

\maketitle

How long does a diffusive particle survive in the presence of absorbing traps? Although deceptively simple, this problem is surprisingly complicated even for static traps \cite{RednerBook,BMS13}. Yet it has wide-ranging applications from diffusion-controlled reactions to animal foraging, search processes and prey-predator dynamics \cite{BLMV11,MOR14}. The problem becomes even harder when the traps are dynamical and even more so if the traps are stochastic. Classical examples include vicious walkers \cite{Fisher84,HF84}, the statistics of the leader (or laggard) among several random-walkers \cite{bA03,BNK10a,BNK10b,HbA18}, and capture processes where a diffusing prey is chased by diffusing predators \cite{BG91,KR96,RK99,GMPR12}. In all of these problems, the survival probability typically decays as a power-law at late-times with a persistence exponent which is notoriously hard to compute. Exact values are known only in a handful of cases \cite{BMS13,DHP95,MSBC96,DHZ96,PS18}.\\

\begin{figure}[t]
\includegraphics[width=\columnwidth]{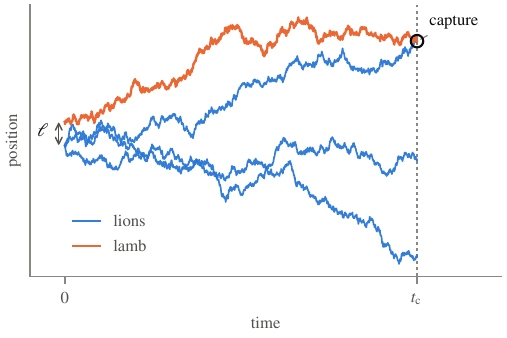}
\caption{Typical trajectories of a lamb (orange) chased by $N = 3$ lions (blue). The lions start together a distance $\ell$ behind the lamb, which survives until the capture time $t_{\rm c}$ at which a lion reaches it for the first time.}
\label{fig:sketch}
\end{figure}

One of the most classical formulations of this problem is known as the lamb and lions problem \cite{BG91,KR96,RK99}. A lamb, starting at $\ell > 0$, and $N$ lions, starting from the origin, perform independent Brownian motions $x_0(t), x_1(t), \cdots, x_N(t)$ with diffusion constant $D$ on a line. Whenever any of the lions captures the lamb the process terminates, see Fig.~\ref{fig:sketch} for example trajectories. The object of interest is the survival probability $S_N(t)$ that the lamb survives up to time $t$ and specifically its late-time decay $S_N(t) \sim t^{-\beta_N}$. Although the lions do not interact they are strongly correlated through their shared termination condition, which is what makes this problem quite hard to solve. For large $N$ the lamb has to stay ahead of the rightmost lion and the problem simplifies to a lamb escaping from a deterministic front closing in like $\sim \sqrt{4 D t \ln N}$ for which $\beta_N \simeq \frac{1}{4} \ln N$ \cite{KR96,RK99,LS02}. Otherwise only two values are known exactly, $\beta_1 = \frac{1}{2}$ and $\beta_2 = \frac{3}{4}$ which are obtained by mapping the problem to a Brownian motion in an $N$-dimensional wedge \cite{FG88,KR96,RK99}.

The first case which has resisted for the past three decades is $N = 3$. It is also a borderline case, it is known that the mean capture time is infinite for $N \leq 3$ but finite for $N \geq 4$ \cite{BG91,LS01,RT09}. So three lions is the largest pride for which the lamb escapes, on average, for an infinite time. The exponent $\beta_3$ has no known closed form solution but has been estimated numerically, first as $\beta_3 \approx 0.91$ \cite{BG91,RK99}, then 0.91342(8) \cite{bA03} and then $0.912\,878\,50$ \cite{HbA18}. To our knowledge, no equation determining $\beta_3$ in closed form is known.

The same number appears in two other fields. In enumerative combinatorics it controls the three-dimensional Kreweras walks, i.e., the lattice walks with steps $(-1,0,0)$, $(0,-1,0)$, $(0,0,-1)$ and $(1,1,1)$ confined to the positive octant \cite{BBKM16,BPRT20}: the number of those which return to the origin after $n$ steps grows as $4^n n^{-(2\beta_3 + 3/2)}$ \cite{DW15,BPRT20,SM}. It is in this context that $\beta_3$ is known most accurately, to 100 digits \cite{DS20}. Its exact value and its rationality are open problems \cite{BPRT20,HJR24}, an irrational value implying that the generating function of these walks is not D-finite \cite{BPRT20,DS20}, and the underlying eigenvalue problem was not expected to be solvable in an explicit manner \cite{HJR24}. In ultracold atoms, $2 \beta_N \hbar \omega$ is the energy of an impurity which repels, with an infinite contact interaction, $N$ noninteracting bosons in a one-dimensional harmonic trap of frequency $\omega$ \cite{SM}. To our knowledge it is known exactly for $N \leq 2$ only \cite{ZVFJV14}, the available exact solutions being restricted to hard walls which sit on all the mirrors of a reflection group \cite{HODVJZ17}, and numerically beyond \cite{DVZ15,RIJ24}.

In this Letter we solve the three lion problem exactly. Let us first summarize our main results. For three lions starting $\ell$ behind the lamb at $t = 0$ the late-time survival probability is given by
\begin{equation} \label{eq:main}
    S_3(t) \sim A \left( \frac{\ell^2}{D t} \right)^{\beta_3}\;, \begin{cases}
        \beta_3 &= 0.912\,878\,502\,087\cdots\\
        A &= 0.154\,612\cdots 
    \end{cases} \;.
\end{equation}
The exponent $\beta_3$ can be expressed in closed form as the smallest root of an explicit transcendental equation (Eq.~\ref{eq:nu-implicit}). We also obtain a closed form for $S_3(t)$ at all times. The core observation that allows us to crack this problem was the identification of a hidden symmetry. The situation is reminiscent of the hydrogen atom, whose
hidden symmetry, associated with the Runge-Lenz vector, is not visible in the geometry of the problem either.

\emph{The model.} We consider a lamb whose position we denote $x_0(t)$ and $N$ lions at positions $x_1(t), \cdots, x_N(t)$ on a line evolving via the Langevin equations
\begin{equation} \label{eq:langevin}
    \begin{dcases}
    \dv{x_i}{t} = \sqrt{2 D}\, \eta_i(t) \\
    x_0(t = 0) = \ell\\
    x_i(t = 0) = 0\;, i \neq 0
    \end{dcases}
    \quad \mbox{~~for~~} i = 0, 1, \cdots, N \;,
\end{equation}
where the $\eta_i(t)$ are independent Gaussian white noises with zero mean and correlator $\langle \eta_i(t) \eta_j(t') \rangle = \delta_{ij} \delta(t - t')$. More general initial conditions are treated in the Supplemental Material \cite{SM}. The lamb survives up to time $t$ with probability $S_N(t)$ if no lion reaches it up to time $t$. Equivalently, a reformulation of the problem is that the leader of $N+1$ walkers remains the leader up to time $t$ \cite{bA03,BNK10b}.

\begin{figure*}[t]
\includegraphics[width=\textwidth]{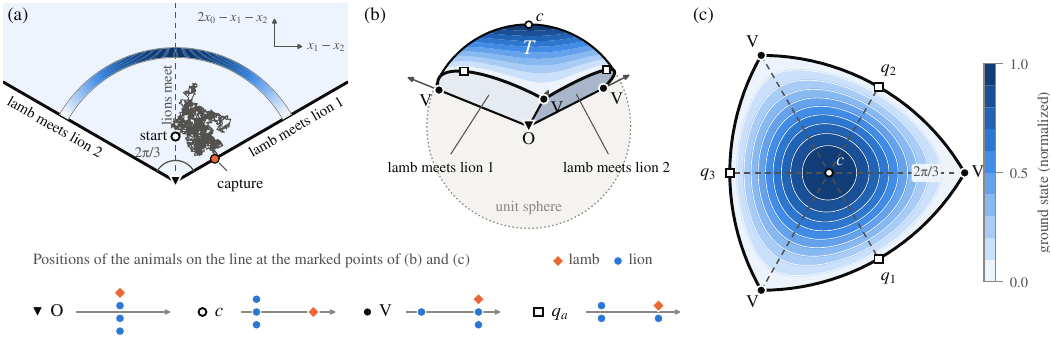}
\caption{Configuration space of the lamb and the lions in the center-of-mass frame. (a) Two lions: the lamb survives as long as a Brownian particle (sample trajectory) stays in a wedge of opening angle $2\pi/3$, whose apex is the point where the three animals coincide. (b) Three lions: the wedge becomes a cone of apex $O$ with three plane walls, on each of which the lamb meets one lion (two of them are visible). The cone cuts the unit sphere along the spherical triangle $T$, which plays the role of the arc of (a). (c) The triangle $T$ in stereographic projection. Solid lines: capture walls. Dashed lines: two lions meet. The walls meet at an angle $2\pi/3$ at the corners $V$, $c$ is the center of $T$ and the squares mark the points where the axes $q_a$ of Eq.~(\ref{eq:def-q}) pierce the sphere, in the middle of the walls. Bottom: positions of the animals on the line at the marked points. Colors: ground state, $\sin(3\phi/2)$ on the arc of (a) and $m_1$ on $T$, which is also the distribution of the configuration of a surviving lamb at late times. An interactive three-dimensional version of panels (b) and (c) is available at \url{https://marcobiroli.github.io/lamb-lions/}.}
\label{fig:config}
\end{figure*}

\emph{Heat equation in a cone.} As is usually the case, it is convenient to consider the survival probability $S_N(\vb{x}, t)$ for arbitrary initial conditions $\vb{x} = (x_0, \cdots, x_N)$. Then writing the backward equation from Eq.~(\ref{eq:langevin}) leads to the well-known heat equation
\begin{equation} \label{eq:heat}
    \pdv{S}{t} = D \sum_{i = 0}^N \pdv[2]{S}{x_i} \;,
\end{equation}
with the initial condition $S = 1$ at $t = 0$. The difficulty in solving Eq.~(\ref{eq:heat}) lies in its boundary conditions. The process dies whenever the lamb touches a lion, i.e. $x_0~\leq~x_i$ for any $i \neq 0$, so $S = 0$ on each of the hyperplanes $x_0 = x_i$ \cite{RednerBook,RK99}. It is common to place oneself in the reference frame of the center of mass then the problem reduces to solving the survival of an $N$-dimensional Brownian motion in a cone, whose apex $O$ is the point where all animals coincide \cite{DeB87,BS97,bA03}. For two lions it is a wedge of opening angle
$2\pi/3$ (see Fig.~\ref{fig:config}a), and for three lions a cone with three plane walls (Fig.~\ref{fig:config}b). 

\emph{Dimensional analysis.} The survival probability is a dimensionless quantity. It can only depend on the position $\vb{r}$ of the particle, measured from the apex of the cone, on the time $t$ and on the diffusion constant $D$ which has the dimension of a length squared over time. Hence the survival probability can only be a function of two dimensionless quantities 
\begin{equation} \label{eq:dimensionless-S}
    S(\alpha, \bm{\omega}) \equiv S\left( \frac{r^2}{4 D t}, \frac{\vb{r}}{r} \right) \;.
\end{equation}
The first parameter $\alpha$ is the ratio of the distance to the diffusive length scale. Early times correspond to $\alpha \to +\infty$ and late times are obtained by taking $\alpha \to 0^+$. The position is then determined by the unit vector $\bm{\omega}$ on the 2-sphere.

\emph{Three lions. } From now on we focus on the three lions $N = 3$ case. In spherical coordinates the Laplacian in Eq.~(\ref{eq:heat}) can be written as
\begin{equation} \label{eq:spherical-laplacian}
    \laplacian = \pdv[2]{}{r} + \frac{2}{r} \pdv{}{r} + \frac{1}{r^2} \laplacian_{\bm{\omega}} \;,
\end{equation}
where $\laplacian_{\bm{\omega}}$ is the Laplacian on the surface of the unit sphere. Note that, up to a sign, it is the usual squared orbital angular momentum operator of quantum mechanics, $\vb{L}^2 = - \laplacian_{\bm{\omega}}$ in units where $\hbar = 1$. Then in the $(\alpha, \bm{\omega})$ coordinates the heat equation in Eq.~(\ref{eq:heat}) becomes \cite{SM}
\begin{equation} \label{eq:spherical-heat}
    \alpha^2 \pdv[2]{S}{\alpha} + \left(\frac{3}{2} + \alpha\right)\alpha \pdv{S}{\alpha} + \frac{1}{4} \laplacian_{\bm{\omega}} S = 0 \;.
\end{equation}
This partial differential equation is linear and $\alpha, \bm{\omega}$ appear in separate terms so we can apply the separation of variables and look for solutions of the form $S(\alpha, \bm{\omega}) \equiv f(\alpha) m(\bm{\omega})$. Inserting this form in Eq.~(\ref{eq:spherical-heat}) yields
\begin{equation} \label{eq:f-m-ODE}
    \frac{1}{f} \left[ \alpha^2 f'' + \left( \frac{3}{2} + \alpha \right) \alpha f' \right] = - \frac{1}{4 m}\laplacian_{\bm{\omega}} m \;.
\end{equation} 
The left-hand-side of Eq.~(\ref{eq:f-m-ODE}) depends only on $\alpha$ whereas the right-hand-side depends only on $\bm{\omega}$ hence they must be both equal to the same constant which we write $\nu(\nu+1)/4$ for reasons which will become apparent soon. Let $T$ denote the spherical triangle which is cut out of the unit 2-sphere in the center-of-mass coordinates by the 3 hyperplanes $x_0 = x_i$ for $i = 1, 2, 3$ (see Fig.~\ref{fig:config}b and~c). We start by looking at the angular part of Eq.~(\ref{eq:f-m-ODE}) which gives
\begin{equation} \label{eq:m-ODE}
    \begin{dcases}
        -\laplacian_{\bm{\omega}} m_j = \nu_j(\nu_j + 1) m_j &\mbox{~~in~~} T \\
        m_j = 0 &\mbox{~~on~~} \partial T 
    \end{dcases} \;.
\end{equation}
This is in fact an eigenvalue problem which corresponds to the Schr\"odinger equation of a free quantum particle that moves on the surface of
the sphere and is confined to $T$ by hard walls, i.e.
a quantum billiard \cite{SS01}. Then $\nu_j$ plays the role of an angular momentum quantum number, and is not an integer because
of the confinement, which is why we chose to write the constant as $\nu(\nu+1)$. In this quantum language, $m_1$ is the ground state and is the only state with constant sign, which we fix to be positive (see Fig.~\ref{fig:config}c). On the other hand, the radial component of Eq.~(\ref{eq:f-m-ODE}) yields
\begin{equation} \label{eq:f-ODE}
    \alpha^2 f_j'' + \left(\frac{3}{2} + \alpha\right)\alpha f_j' - \frac{1}{4} \nu_j(\nu_j + 1) f_j = 0 \;. 
\end{equation}
To solve it we start by considering the small $\alpha$ limit (i.e. the long-time limit) where we expect $f_j(\alpha) \sim \alpha^s$ and we can approximate $3/2 + \alpha \sim 3/2$ in Eq.~(\ref{eq:f-ODE}) which gives
\begin{equation} \label{eq:s}
    s(s+1/2) = \frac{1}{4} \nu_j(\nu_j + 1)  \;. 
\end{equation}
Hence $s = \nu_j / 2$ or $s = - (\nu_j + 1)/2$ but the second solution is divergent for $\alpha \to 0^+$ which is not possible for a probability. Therefore we take $f(\alpha) = \alpha^{\nu_j/2} g(\alpha)$ which when plugged back into Eq.~(\ref{eq:f-ODE}) gives
\begin{equation} \label{eq:g-ODE}
    \alpha g'' + \left(\nu_j + \frac{3}{2} + \alpha\right) g' + \frac{1}{2} \nu_j g = 0 \;.
\end{equation}
With $z = -\alpha$ this is Kummer's equation \cite{DLMF} hence
\begin{equation} \label{eq:f-solution}
    f_j(\alpha) = \alpha^{\nu_j/2} \,_1F_1\left(\frac{\nu_j}{2}; \nu_j + \frac{3}{2} ; - \alpha\right) \;,
\end{equation}
where $\,_1F_1(a;b;z)$ is the confluent hypergeometric function. The asymptotics of the radial function can be obtained from the known asymptotics of the hypergeometric function
\begin{equation}\label{eq:f-asymptotic}
    f_j(\alpha) \to \begin{dcases}
        \frac{\alpha ^{\nu_j /2} (3-(\alpha -2) \nu_j )}{2 \nu_j +3} &\mbox{~~when~~} \alpha \ll 1\\
        \frac{\Gamma \left(\nu_j +\frac{3}{2}\right)}{\Gamma \left(\frac{\nu_j +3}{2}\right)} &\mbox{~~when~~} \alpha \to +\infty
    \end{dcases} \;.
\end{equation}
Coming back to the original Eq.~(\ref{eq:spherical-heat}), its linearity imposes that the solution has to be a linear superposition $S = \sum_j B_j m_j(\bm{\omega}) f_j(\alpha)$. From Eq.~(\ref{eq:m-ODE}) we know that $m_j(\bm{\omega})$ are orthonormal and using the early-time limit of $f_j(\alpha)$ in Eq.~(\ref{eq:f-asymptotic}) together with the boundary condition $S = 1$ at $t = 0$ we get
\begin{equation} \label{eq:Bj-solution}
    B_j = \frac{\Gamma(\frac{\nu_j + 3}{2})}{\Gamma(\nu_j + 3/2)} \int \dd \bm{\omega} \; m_j(\bm{\omega}) \;.
\end{equation}
When the three lions start together at a distance $\ell$ behind the lamb, the particle starts on the axis of the cone, in the direction $c$ of the center of $T$, at a distance $r = \sqrt{3} \ell/2$ from the apex. We thus obtain \cite{SM}
\begin{equation} \label{eq:S3-series}
S_3(t) = \sum_{j \geq 1} A_j \alpha^{\nu_j/2} \,_1F_1(\frac{\nu_j}{2}; \nu_j + \frac{3}{2}; -\alpha) \;,
\end{equation}
with $A_j= B_j m_j(c)$ and $\alpha = 3 \ell^2/16 D t$. This expression is exact at all times,
but it is only implicit: it requires the eigenvalues $\nu_j$ and
the eigenfunctions $m_j$ of Eq.~(\ref{eq:m-ODE}). At late times, $\alpha \to 0$, the $j$-th term decays as $t^{-\nu_j /2}$. The ground state therefore dominates, $S_3(t) \simeq A_1 \alpha^{\nu_1 /2}$, and
\begin{equation} \label{eq:beta-nu}
    \beta_3 = \frac{\nu_1}{2}\;.
\end{equation}
Solving the three lion problem therefore reduces to finding the ground state energy of the Schr\"odinger equation in Eq.~(\ref{eq:m-ODE}) \cite{RT09}. For two lions the same steps apply in two dimensions. The modes are $\sin(\mu \phi)$ on the arc of Fig.~\ref{fig:config}a, with $\mu = 3/2,3,\cdots$, the $\mu$-th term
decays as $t^{-\mu/2}$, and we recover $\beta_2 = 3/4$.

\emph{A hidden symmetry.}  The eigenvalue problem of Eq.~(\ref{eq:m-ODE}), could so far be solved only numerically. The key to its solution, and the main contribution of this Letter, is a conservation law, which is easiest to describe classically. We place ourselves in the following coordinates
\begin{align}
    q_1 = (x_0 + x_1 - x_2 - x_3)/2 \;,\nonumber\\
    q_2 = (x_0 - x_1 + x_2 - x_3)/2 \;,\label{eq:def-q}\\
    q_3 = (x_0 - x_1 - x_2 + x_3)/2 \;.\nonumber
\end{align}
The three capture events are defined by the walls $q_j + q_k = 0$ while the planes $q_j - q_k = 0$ are symmetry planes corresponding to when two lions cross paths. Together these six planes are the diagonal mirror planes of a cube with axes $q_1, q_2, q_3$, and the cone is the region of space
closest to one vertex of a regular tetrahedron inscribed in this cube.

\begin{figure}[t]
\includegraphics[width=\columnwidth]{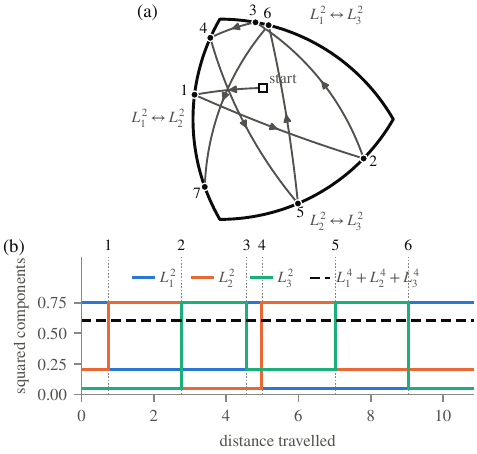}
\caption{Billiard picture of the hidden conserved quantity. (a) A ball moves along great circles inside the triangle $T$ of Fig.~\ref{fig:config}c and bounces off the capture walls, the numbers label the successive bounces. (b) Squared components of its angular momentum along the axes $q_1, q_2, q_3$ as a function of the distance travelled. Each bounce exchanges two of them, as indicated next to the walls in (a), so that $L_1^4 + L_2^4 + L_3^4$ (dashed line) is conserved.}
\label{fig:billiard}
\end{figure}

Consider now the classical counterpart of this problem. A ball moves freely inside the cone and bounces elastically off its walls (see Fig.~\ref{fig:billiard}a). Between two bounces its angular momentum $\vb{L} = (L_1,L_2,L_3)$ is conserved. Since every wall is a mirror plane of the cube, a bounce merely
exchanges two components of $L$, up to signs. Any symmetric function of $L^2_1$, $L^2_2$ and $L^2_3$ is therefore conserved forever. The simplest one, $L^2_1 + L^2_2 + L^2_3$, is the energy. The next one,
\begin{equation}\label{eq:def-O}
    \mathcal{O}_4 = L^4_1 + L^4_2 + L^4_3 \;,
\end{equation}
is a new constant of motion (see Fig.~\ref{fig:billiard}b). Hence, the classical billiard is integrable. In fact, all its trajectories are periodic \cite{SS01}.

\begin{figure}[t]
\includegraphics[width=\columnwidth]{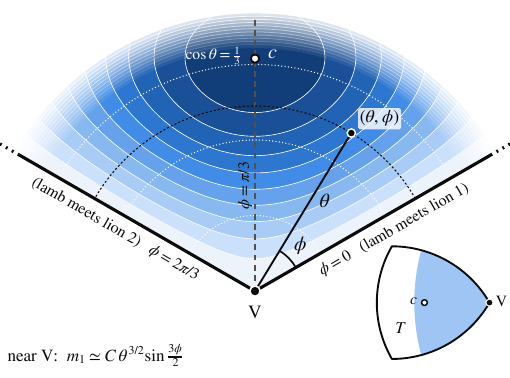}
\caption{Zoom on a corner $V$ of the triangle $T$ (shaded region of the inset) in the corner coordinates $(\theta, \phi)$: $\theta$ is the angular distance to $V$ and $\phi$ is the angle around it, measured from the wall on which the lamb meets lion 1. The two capture walls which meet at $V$ are the lines $\phi = 0$ and $\phi = 2\pi/3$, and the two lions meet on the dashed line $\phi = \pi/3$, which goes through the center $c$ of $T$ located at $\cos \theta = 1/3$. Dotted lines: circles of constant $\theta$. Colors: ground state $m_1$, with the same color scale as in Fig.~\ref{fig:config}c. Close to the corner it reduces to the two-lion form of Eq.~(\ref{eq:ground-vertex}).}
\label{fig:corner}
\end{figure}

Does this conservation law survive in its quantum counterpart? There $\vb{L} = - i \vb{q} \times \nabla$ is the angular momentum operator, and the Hamiltonian is $\vb{L}^2 = - \laplacian_{\bm{\omega}}$. Conservation in the quantum case means that the symmetry operator commutes with the Hamiltonian $[\hat{\mathcal{O}}_4, \vb{L}^2] = 0$. Away from the corners this is guaranteed by the mirror symmetries. At the corners, however, waves diffract, and in planar billiards whose angles are not of the form $\pi/n$ the classical conserved quantities are known not to survive in the quantum problem \cite{RB81,KGW24}, for the same reason the method of images fails. Three copies of $T$ surround a corner, and alternating signs cannot be assigned consistently to an odd number of them. Physically, being close to a corner corresponds to a situation where the lamb is being approached by two lions while the third is still far away. Hence to first order the neighborhood of the corner should behave like the two lion problem which has a known solution
\begin{equation} \label{eq:ground-vertex}
    m_1 \simeq C \theta^{3/2} \sin \frac{3 \phi}{2}, \mbox{~~as~~} \theta \to 0^+ \;,
\end{equation}
where $\theta$ is the angular distance to the corner and $\phi$ is the angle around it (see Fig.~\ref{fig:corner}). Now acting on $m_1$ with $\hat{\mathcal{O}}_4$ would normally generate a singular divergence in $\theta^{-3/2}$ but for a corner of angle $2 \pi / 3$ this divergence vanishes (see the Supplemental Material for a detailed derivation \cite{SM}). Consequently, the eigenstates can be chosen such that
\begin{equation} \label{eq:O4-eigen}
    \hat{\mathcal{O}}_4 m_j = \kappa_j m_j \;.
\end{equation}
Each level has two quantum numbers, $\nu_j$ and $\kappa_j$, which makes this system with two degrees of freedom integrable. The situation is reminiscent of the hydrogen atom, whose hidden symmetry, associated with the Runge-Lenz vector, is not visible in the geometry of the problem either. The operator of Eq.~(\ref{eq:def-O}) is in fact familiar: it is the cubic crystal-field Hamiltonian of a magnetic ion, and it splits the rotational levels of spherical-top molecules \cite{LLW62,AB70,Hecht61}. Solving the three-lion problem is thus equivalent to diagonalizing a cubic crystal field, with a noninteger angular momentum $\nu$. Conserved quantities of this kind, i.e., invariant polynomials of a reflection group evaluated on the angular momenta, are known for the angular Calogero model \cite{CL15} and for hard-core particles in a harmonic trap \cite{HODVJZ17}. There, however, the particle is confined to a single chamber of the group: all the mirrors are walls and all the corners are of the form $\pi/n$. Here only three of the six mirrors are walls, the triangle $T$ is the union of six chambers, and the conservation law holds only thanks to the cancellation at its corners.

\emph{Exact solution.} To solve the problem exactly we place ourselves in the same corner coordinates $(\theta, \phi)$ as in Eq.~(\ref{eq:ground-vertex}) (see Fig.~\ref{fig:corner}). Then writing the ground state as $m_1(\theta, \phi) \equiv g(\theta) \sin(\mu \phi)$ we get 
\begin{equation} \label{eq:ground-legendre}
    m_1(\theta, \phi) = \sum_{k \geq 0} c_k P_\nu^{-\mu_k}(\cos \theta) \sin(\mu_k \phi) \;,
\end{equation}
with $\mu_k = 3k + 3/2$ (see the Supplemental Material). The symmetry operator $\mathcal{O}_4$ in this basis is 
\begin{equation} \label{eq:O-matrix}
    \mathcal{O} = \begin{pmatrix}
        D_0 & -\sqrt{2} & & \\
        -\sqrt{2} \Lambda_0 & D_1 & -2 \sqrt{2} & \\
         & -2 \sqrt{2}\Lambda_1 & D_2 & \ddots \\
         & & \ddots & \ddots 
    \end{pmatrix} \;,
\end{equation}
where $D_k = -\frac{7}{6}\mu_k^4 + (\lambda - 5/6)\mu_k^2 + \lambda^2/2$, with $\lambda = \nu(\nu + 1)$ and $\Lambda_k = \prod_{i = 0}^{2} (\nu - \mu_k - i)(\nu + \mu_k + 1 + i)$. Then Eq.~(\ref{eq:O4-eigen}) means that the coefficients $c_k$ in Eq.~(\ref{eq:ground-legendre}) are eigenvectors of the tri-diagonal matrix in Eq.~(\ref{eq:O-matrix}) with eigenvalue $\kappa$. This hidden conserved quantity has reduced the two-dimensional problem on the sphere to a one-dimensional one. 

To diagonalize $\mathcal{O}$ we can exploit the tri-diagonal structure in Eq.~(\ref{eq:O-matrix}). The tri-diagonal structure gives us a recursive equation for the ratio $\rho_k = c_k / c_{k-1}$ of consecutive coefficients
\begin{equation} \label{eq:rhok-rhokp1}
    \rho_k = \frac{\sqrt{2} k \Lambda_{k - 1}}{D_k - \kappa - \sqrt{2}(k + 1) \rho_{k + 1}} \;.
\end{equation}
The row $k = 0$ initializes the recursion with $\kappa = D_0 - \sqrt{2} \rho_1$ and applying Eq.~(\ref{eq:rhok-rhokp1}) recursively gives $\kappa$ in a closed form as an infinite continued fraction
\begin{equation}\label{eq:kappa}
    \kappa(\nu) = D_0 - \cfrac{2 \Lambda_0}{D_1 - \kappa - \cfrac{8 \Lambda_1}{D_2 - \kappa - \cfrac{18 \Lambda_2}{D_3 - \kappa - \cdots}}} \;.
\end{equation}
For a fixed $\nu$ this equation can be solved easily by iteration. We start with $\kappa = D_0$ and plug it in the right hand side and iterate until a fixed point. In practice this sequence converges extremely quickly and truncating the infinite fraction to $K$ rows introduces an error of order $8^{-K}$ \cite{SM}. Then the coefficients $c_k(\nu)$ can be obtained via
\begin{equation} \label{eq:ck}
    c_k(\nu) = \rho_1 \rho_2 \cdots \rho_k \;.
\end{equation}

Now that $\kappa$ is fixed, we need to fix $\nu$. To do so we will use the symmetry at the center of the triangle ($c$ in Figs.~\ref{fig:config}c and~\ref{fig:corner}). At the center the three lions are superposed and hence should be exchangeable. This is the safest configuration for the lamb since it maximizes the distance to any of the lions. Hence $m_1$ should be maximal in $c$ which gives us a condition on the gradient of $m_1$ which has to cancel
\begin{equation} \label{eq:nu-implicit}
    \grad m_1 |_{c} = \sum_{k \geq 0} (-1)^k c_k(\nu) \dv{}{\theta} P_\nu^{-\mu_k} (\cos \theta) \Big|_{\cos \theta = 1/3} = 0 \;.
\end{equation}
The smallest positive root $\nu_1$ of Eq.~(\ref{eq:nu-implicit}) gives the three-lion exponent $\beta_3 = \nu_1/2$ thanks to Eq.~(\ref{eq:beta-nu}). In practice, $\nu$ can be obtained to arbitrary accuracy by a root search on Eq.~(\ref{eq:nu-implicit}). Doing so the first digits of $\beta_3$ are
\begin{equation}
    \beta_3 = 0.912\,878\,502\,087\,228\,125\,48\cdots \;.
\end{equation}

\begin{figure}[t]
\includegraphics[width=\columnwidth]{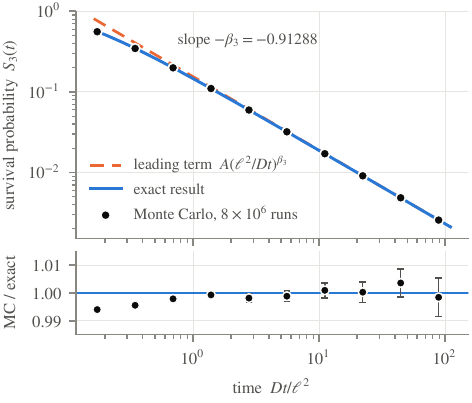}
\caption{Survival probability $S_3(t)$ of the lamb for three lions starting together a distance $\ell$ behind it. Symbols: numerical simulations ($8 \times 10^6$ realizations of lattice random walks with $\ell = 12$ lattice spacings). Solid line: exact result of Eq.~(\ref{eq:S3-series}). Dashed line: late-time behavior of Eq.~(\ref{eq:main}). Lower panel: ratio of the simulations to the exact result, the small deviation at early times is due to the lattice.}
\label{fig:survival}
\end{figure}

The same construction gives the excited states which are symmetric in the lions, each with its own pair $(\nu_j, \kappa_j)$, and hence all the non-vanishing terms of Eq.~(\ref{eq:S3-series}) \cite{SM}. In Fig.~\ref{fig:survival} we compare it with numerical simulations and find an excellent agreement at all times, without any fitting parameter.

\emph{Ground state.} The ground state series expansion in Eq.~(\ref{eq:ground-legendre}) converges very quickly so the first term is a good approximation (within 0.3\% of the full expression) of the ground state and is explicitly given by
\begin{align}
    m_1 &\simeq P_\nu^{-3/2}(\cos \theta) \sin(3\phi/2) \nonumber\\
    &= \sqrt{\frac{2}{\pi}} \frac{\cot \theta \sin a \theta - a \cos a\theta}{a(a^2 - 1)\sqrt{\sin \theta}} \sin \frac{3 \phi}{2} \;, \label{eq:ground-first}
\end{align}
where $a = \nu_1 + 1/2$. Eq.~(\ref{eq:ground-first}) has a simple interpretation. It is almost exactly the two-lion solution of
Eq.~(\ref{eq:ground-vertex}), dressed by the curvature of the sphere. In other words, seen from the nearest corner, the third lion contributes almost nothing to the ground state. It makes itself felt essentially only through the eigenvalue of the ground state $\nu_1$ (and hence $\beta_3$).

The ground state itself also has a physical meaning. For a lamb that is still alive at a late time $t \gg \ell^2/D$, the four animals are spread over a distance of order $\sqrt{D t}$, and the shape of their configuration is distributed over $T$ with a density proportional to $m_1(\bm{\omega})$ (see Fig.~\ref{fig:config}c) \cite{SM}. For instance, a surviving lamb at time $t$ leads the nearest lion by $1.918\sqrt{D t}$ on average, which we have checked numerically \cite{SM}. As expected this is below the lead of a lamb chased by a single lion, $\sqrt{2 \pi D t} \simeq 2.51 \sqrt{D t}$, or by two lions, $2.09 \sqrt{D t}$ \cite{SM}, meaning that the larger the pride the closer it presses the lamb.

\emph{Other orderings and more lions.} The mechanism is not specific to the lamb and the lions. It only requires that the walls be mirror planes of the cube and that every corner be either of the form $\pi/n$ or a protected one, of angle $2\pi/3$. This is the case for several ordering problems of four random walkers~\cite{bA03,HbA18}. We have verified Eq.~\eqref{eq:O4-eigen} for three of them, and our exponents agree with the numerical values of Ref.~\cite{HbA18} (see Table~\ref{tab:orderings}). Two of them are capture problems: two lambs chased by two lions (second line), and a lamb caught between one lion ahead of it and two behind (fourth line). In contrast, the laggard problem~\cite{bA03}, i.e., the probability that a given walker never takes the lead, has corners of angle $4\pi/3$, for which the dangerous term does not vanish.

\begin{table}[t]
\caption{Ordering problems of four random walkers $A$, $B$, $C$ and $D$ on a line solved by Eq.~\eqref{eq:O4-eigen}: the probability that the ordering of the first column is maintained up to time $t$ decays as $t^{-\beta}$. Here $A > B$ means that $A$ stays ahead of $B$, and the relative order of the walkers inside braces is free. First line: $A$ remains the leader, i.e., a lamb and three lions. Second line: $A$ and $B$ both stay ahead of $C$ and $D$. Third line: $A$ remains the leader and $D$ the laggard. Fourth line: $A$ remains first and $B$ second. Last column: numerical values of Ref.~\cite{HbA18}, where these four problems are called leader, teams, bookends and $12{\bullet}{\bullet}$.}
\label{tab:orderings}
\begin{ruledtabular}
\begin{tabular}{lll}
Ordering maintained & \multicolumn{1}{c}{$\beta$ (exact)} & Ref.~\cite{HbA18}\\
\hline
$A > \{B, C, D\}$ & 0.9128785021 & 0.91287850\\
$\{A, B\} > \{C, D\}$ & 1.1949399623 & 1.1949400\\
$A > \{B, C\} > D$ & 1.8737525413 & 1.8737525\\
$A > B > \{C, D\}$ & 2.0716054591 & 2.0716054\\
\end{tabular}
\end{ruledtabular}
\end{table}

For four lions the billiard still has conserved quantities of the type of Eq.~\eqref{eq:def-O}, but an exhaustive search up to tenth order in the angular momenta shows that none of them survives diffraction~\cite{SM}. The three-lion solution can nevertheless serve as the first step of a recursive construction of the $N$-lion states, described in the Supplemental Material~\cite{SM}, which yields $\beta_4=1.032\,267\,875$ and $\beta_5=1.125\,865\,77$. In particular $\beta_3<1<\beta_4$, consistent with a finite mean capture time for four lions~\cite{RT09}. The construction also provides an exact relation between $\beta_{N-1}$ and $\beta_N$. For a large pride it indicates that each additional lion raises the exponent by $1/(4N)$, which accounts for the logarithmic law $\beta_N\simeq\frac14\ln N$~\cite{KR96,LS02,SM}.

\emph{Conclusion.} To summarize, we have solved exactly the capture of a diffusing lamb by three diffusing lions, the simplest capture process beyond the reach of the method of images. The solution rests on a hidden conserved quantity, which survives diffraction at the corners of the absorbing cone precisely because of their special angle $2\pi/3$, and which reduces the problem to that of a single spin in a cubic crystal field. It also gives the critical exponent of the three-dimensional Kreweras walks as the root of an explicit equation, while the question of its rationality~\cite{BPRT20,HJR24} remains open. A natural question concerns animals with different diffusion constants~\cite{KR96,RK99}, for which the walls are no longer mirror planes of the cube. We hope that the mechanism uncovered here will be useful for other first-passage problems of interacting or ordered particles and, through the correspondence with trapped impurities \cite{SM,ZVFJV14,HODVJZ17,RIJ24}, for few-body problems of ultracold atoms.

\emph{Code availability.} The code which implements the solution and produces the figures, together with the data of the simulations, is openly available at \url{https://github.com/MarcoBiroli/lamb_lions}.

\begin{acknowledgments}
    {\bf Acknowledgments.} This research received support through Schmidt Sciences, LLC. The calculations and the simulations of this manuscript were prepared with the assistance of Claude (Anthropic). The Supplemental Material was drafted by Claude (Anthropic) and is currently being cleaned-up/reviewed by the author.
\end{acknowledgments}

\bibliography{refs}

\input{supplement}

\end{document}

%% file: supplement.tex
\clearpage
\onecolumngrid
\setcounter{equation}{0}
\setcounter{figure}{0}
\setcounter{table}{0}
\setcounter{section}{0}
\setcounter{secnumdepth}{1}
\renewcommand{\thesection}{\Roman{section}}
\renewcommand{\theequation}{S\arabic{equation}}
\renewcommand{\thefigure}{S\arabic{figure}}
\renewcommand{\thetable}{S\arabic{table}}
\makeatletter
\renewcommand{\theHequation}{S\arabic{equation}}
\renewcommand{\theHfigure}{S\arabic{figure}}
\renewcommand{\theHtable}{S\arabic{table}}
\makeatother

\begin{center}
\textbf{\large Supplemental Material for ``Capture of a diffusing lamb by three diffusing lions:\\ hidden symmetry and an exact solution''}
\end{center}

In this Supplemental Material we give the details of the calculations described in the main text. Equations and tables of the Supplemental Material carry the prefix S, all the others refer to the main text. It is organized as follows. In Sec.~\ref{S:sec-cone} we derive the heat equation in the cone for $N$ lions, treat general initial conditions and describe the configuration of a surviving lamb. In Sec.~\ref{S:sec-conserved} we show that the conserved quantity of Eq.~(\ref{eq:def-O}) survives at the corners, which leads to Eq.~(\ref{eq:O4-eigen}). In Sec.~\ref{S:sec-matrix} we derive the tri-diagonal matrix of Eq.~(\ref{eq:O-matrix}) and the continued fraction of Eq.~(\ref{eq:kappa}). In Sec.~\ref{S:sec-center} we derive the condition at the center, Eq.~(\ref{eq:nu-implicit}), and give the ground state explicitly. Sections~\ref{S:sec-checks} and~\ref{S:sec-simulations} describe independent checks and the simulations of Fig.~\ref{fig:survival}. Section~\ref{S:sec-orderings} deals with the other orderings of Table~\ref{tab:orderings} and with four lions, Sec.~\ref{S:sec-recursion} with the recursive construction for $N$ lions and Sec.~\ref{S:sec-impurity} with the correspondence with an impurity in a trapped Bose gas.

\section{From the lamb and the lions to a Brownian particle in a cone} \label{S:sec-cone}

\emph{The cone.} The vector $\vb{x} = (x_0, x_1, x_2, x_3)$ performs an isotropic Brownian motion in four dimensions. The center of mass $\bar{x} = \frac{1}{4} \sum_i x_i$ plays no role in the capture, so we are left with three dimensions for which we use the orthonormal coordinates $(q_1, q_2, q_3)$ of Eq.~(\ref{eq:def-q}). Since
\begin{equation} \label{S:walls}
    q_2 + q_3 = x_0 - x_1 \;, \quad q_1 + q_3 = x_0 - x_2 \;, \quad q_1 + q_2 = x_0 - x_3 \;,
\end{equation}
the lamb survives as long as the particle $\vb{r} = (q_1, q_2, q_3)$ stays in the cone where the three combinations of Eq.~(\ref{S:walls}) are positive. Note that $r^2 = \sum_{i} (x_i - \bar{x})^2$. The cone cuts the unit sphere along the spherical triangle $T$ of Fig.~\ref{fig:config}. Its center is $c = (1, 1, 1)/\sqrt{3}$, its three corners are $V = (1, 1, -1)/\sqrt{3}$ and its permutations, its three angles are equal to $2\pi/3$ and its area is $\pi$, i.e. a quarter of the sphere. The axes $q_a$ cross its sides at their midpoints.

\emph{Heat equation in the $(\alpha, \bm{\omega})$ coordinates.} We give the derivation of Eq.~(\ref{eq:spherical-heat}) directly for $N$ lions, i.e. in $N$ dimensions, since it comes at no extra cost. For a function of $\alpha = r^2/4Dt$ and of the direction $\bm{\omega}$ we have
\begin{equation} \label{S:chain-rule}
    \pdv{}{t} = - \frac{\alpha}{t} \pdv{}{\alpha} \;, \quad \pdv{}{r} = \frac{2 \alpha}{r} \pdv{}{\alpha} \;, \quad \pdv[2]{}{r} = \frac{4 \alpha^2}{r^2} \pdv[2]{}{\alpha} + \frac{2 \alpha}{r^2} \pdv{}{\alpha} \;.
\end{equation}
Plugging these in the $N$-dimensional Laplacian $\laplacian = \partial_r^2 + \frac{N-1}{r} \partial_r + \frac{1}{r^2} \laplacian_{\bm{\omega}}$, the heat equation in Eq.~(\ref{eq:heat}) becomes
\begin{equation} \label{S:heat-N}
    \alpha^2 \pdv[2]{S}{\alpha} + \left( \frac{N}{2} + \alpha \right) \alpha \pdv{S}{\alpha} + \frac{1}{4} \laplacian_{\bm{\omega}} S = 0 \;,
\end{equation}
which is Eq.~(\ref{eq:spherical-heat}) for $N = 3$. We then look for solutions of the form $f(\alpha) m(\bm{\omega})$ with $- \laplacian_{\bm{\omega}} m = \nu (\nu + N - 2) m$. The same steps as in the main text give $f = \alpha^{\nu/2} g$ where $g$ satisfies Kummer's equation, hence
\begin{equation} \label{S:f-N}
    \alpha g'' + \left( \nu + \frac{N}{2} + \alpha \right) g' + \frac{\nu}{2} g = 0 \quad \Longrightarrow \quad f(\alpha) = \alpha^{\nu/2} \,_1F_1\left( \frac{\nu}{2} ; \nu + \frac{N}{2} ; - \alpha \right) \;,
\end{equation}
which gives back Eq.~(\ref{eq:f-solution}) for $N = 3$ \cite{DLMF}. The second solution of Kummer's equation corresponds to the root $s = - (\nu + N - 2)/2$ which we discarded in the main text. As a sanity check, for a single lion ($N = 1$, $\nu = 1$ and $\alpha = \ell^2/8Dt$) these formulas give back the textbook result $S_1(t) = \frac{2}{\sqrt{\pi}} \alpha^{1/2} \,_1F_1(\frac{1}{2}; \frac{3}{2}; -\alpha) = \mathrm{erf} \sqrt{\alpha}$.

\emph{General initial conditions.} Let $m_j$ be the modes of Eq.~(\ref{eq:m-ODE}), normalized as $\int_T m_j^2 \, \dd \bm{\omega} = 1$. Repeating the steps of the main text for a particle which starts at a distance $r_0$ from the apex of the cone, in the direction $\bm{\omega}_0$, we get \cite{DeB87,BS97}
\begin{equation} \label{S:S-general}
    S_3(t) = \sum_{j \geq 1} B_j \, m_j(\bm{\omega}_0) \, \alpha^{\nu_j/2} \,_1F_1\left( \frac{\nu_j}{2} ; \nu_j + \frac{3}{2} ; -\alpha \right) \;, \mbox{~~where~~} \alpha = \frac{r_0^2}{4 D t} \;,
\end{equation}
with the coefficients $B_j$ of Eq.~(\ref{eq:Bj-solution}). In terms of the animals $r_0^2 = \sum_{i = 0}^{3} [x_i(0) - \bar{x}(0)]^2$ and $\bm{\omega}_0$ is the direction of the initial configuration. For three lions starting together at a distance $\ell$ behind the lamb we have $\bm{\omega}_0 = c$ and $r_0^2 = 3 \ell^2/4$, which gives Eq.~(\ref{eq:S3-series}) with $A_j = B_j m_j(c)$. Note that in this case only the modes which are symmetric under the exchange of the three lions contribute. At late times the ground state dominates for any initial condition,
\begin{equation} \label{S:S-late}
    S_3(t) \simeq B_1 \, m_1(\bm{\omega}_0) \left( \frac{r_0^2}{4 D t} \right)^{\beta_3} \;,
\end{equation}
so the initial condition only enters through the prefactor. Expanding the first term of Eq.~(\ref{eq:S3-series}) for small $\alpha$ with the help of Eq.~(\ref{eq:f-asymptotic}) also gives the leading correction to the power law, $1 - \nu_1 \alpha/(2 \nu_1 + 3) = 1 - 0.0515 \, \ell^2/Dt$. The second mode only contributes at relative order $\alpha^{(\nu_2 - \nu_1)/2} \simeq \alpha^{1.45}$.

\emph{Configuration of a surviving lamb.} The same modes give the propagator of the particle in the cone
\begin{equation} \label{S:propagator}
    G(r, \bm{\omega}, t \,|\, r_0, \bm{\omega}_0) = \frac{e^{-(r^2 + r_0^2)/4Dt}}{2 D t \sqrt{r \, r_0}} \sum_{j \geq 1} m_j(\bm{\omega}) \, m_j(\bm{\omega}_0) \, I_{\nu_j + \frac{1}{2}}\left( \frac{r \, r_0}{2 D t} \right) \;,
\end{equation}
where $I_\sigma(z)$ is the modified Bessel function. Integrating it over the cone gives back Eq.~(\ref{S:S-general}). For $D t \gg r_0^2$ the first term dominates and $I_{\nu_1 + 1/2}(z) \propto z^{\nu_1 + 1/2}$. Hence the position of a particle which is still alive at time $t$ is distributed, independently of the initial condition, with the density
\begin{equation} \label{S:conditional}
    P(r, \bm{\omega} \,|\, t_{\rm c} > t) \simeq \frac{r^{\nu_1} \, e^{-r^2/4Dt}}{2^{\nu_1 + 2} \, \Gamma\left( \frac{\nu_1 + 3}{2} \right) (D t)^{(\nu_1 + 3)/2}} \; \frac{m_1(\bm{\omega})}{\int_T m_1 \, \dd \bm{\omega}} \;,
\end{equation}
with respect to $r^2 \dd r \, \dd \bm{\omega}$. The size $r$ and the shape $\bm{\omega}$ of the configuration are therefore independent. The shape is distributed over $T$ with a density proportional to $m_1(\bm{\omega})$, as stated in the main text, and $\avg{r^2} = 2 (\nu_1 + 3) D t \simeq 9.6515 \, D t$, to be compared with $6 D t$ for four free particles. The lead of the lamb over the nearest lion is $\sqrt{2}$ times the distance of the particle to the nearest wall, i.e. $r$ times a function of $\bm{\omega}$. Averaging it with Eq.~(\ref{S:conditional}) gives
\begin{equation} \label{S:lead}
    \avg{x_0 - \max_{i \geq 1} x_i} \simeq 1.91837 \sqrt{D t} \;, \quad \avg{\max_{i \geq 1} x_i - \min_{i \geq 1} x_i} \simeq 1.94505 \sqrt{D t} \;,
\end{equation}
for the lead of a surviving lamb and for the extension of the pride which chases it. For one and two lions the same calculation is elementary, since the ground states are $1$ and $\sin(3\phi/2)$, and gives a lead of
\begin{equation} \label{S:lead-12}
    \sqrt{2 \pi D t} \simeq 2.5066 \sqrt{D t} \mbox{~~for~~} N = 1 \;, \quad \frac{6 \sqrt{2}}{5} \frac{\Gamma(9/4)}{\Gamma(7/4)} \sqrt{D t} \simeq 2.0921 \sqrt{D t} \mbox{~~for~~} N = 2 \;.
\end{equation}
The larger the pride the closer it presses the lamb. We checked Eq.~(\ref{S:lead}) with simulations of $6 \times 10^7$ realizations of the lattice model of Sec.~\ref{S:sec-simulations} with $\ell = 4$. At $D t/\ell^2 = 200$ they give a mean lead of $1.920(4) \sqrt{D t}$ and $\avg{r^2} = 9.69(2) \, D t$. For one and two lions the same simulations ($4 \times 10^7$ realizations with $\ell = 2$, at $D t/\ell^2 = 125$) give a mean lead of $2.507(1) \sqrt{D t}$ and $2.091(2) \sqrt{D t}$, in agreement with Eq.~(\ref{S:lead-12}).

\section{The hidden conserved quantity} \label{S:sec-conserved}

In this section we show that the classical constant of motion of Eq.~(\ref{eq:def-O}) survives in the quantum problem, i.e. that the modes can be chosen to satisfy Eq.~(\ref{eq:O4-eigen}). With $\vb{L} = - i \vb{q} \times \nabla$ the operator reads explicitly
\begin{equation} \label{S:O4-explicit}
    \hat{\mathcal{O}}_4 = \left( q_2 \partial_3 - q_3 \partial_2 \right)^4 + \left( q_3 \partial_1 - q_1 \partial_3 \right)^4 + \left( q_1 \partial_2 - q_2 \partial_1 \right)^4 \;,
\end{equation}
where $\partial_j = \partial/\partial q_j$. There is no ordering ambiguity since each term is a power of a single component of $\vb{L}$. For $\hat{\mathcal{O}}_4$ to be conserved it has to map every solution $m$ of Eq.~(\ref{eq:m-ODE}) onto another solution with the same $\nu$. Three conditions must be met: (i) $\hat{\mathcal{O}}_4 m$ obeys the same differential equation as $m$, (ii) $\hat{\mathcal{O}}_4 m$ vanishes on the walls and (iii) $\hat{\mathcal{O}}_4 m$ remains finite at the corners, where two walls meet. We go through them one by one.

\emph{(i) The differential equation.} The components of $\vb{L}$ generate the rotations of the sphere, which leave its Laplacian unchanged. Explicitly $[L_j, L_k] = i \sum_l \epsilon_{jkl} L_l$, hence $[L_j, \vb{L}^2] = i \sum_{k,l} \epsilon_{jkl} (L_l L_k + L_k L_l) = 0$ and therefore $[\hat{\mathcal{O}}_4, \vb{L}^2] = 0$.

\emph{(ii) The walls.} The reflection $R$ in the wall $q_2 + q_3 = 0$ is $(q_1, q_2, q_3) \to (q_1, -q_3, -q_2)$. Plugging it in $\vb{L} = - i \vb{q} \times \nabla$ we see that it maps $(L_1, L_2, L_3)$ to $(-L_1, L_3, L_2)$, and similarly for the two other walls. A reflection merely exchanges two components of $\vb{L}$ up to signs, which is the quantum counterpart of the bounces of Fig.~\ref{fig:billiard}, and therefore $\hat{\mathcal{O}}_4$ commutes with $R$. More generally $\hat{\mathcal{O}}_4$ is invariant under all the signed permutations of $(q_1, q_2, q_3)$, i.e. under the symmetries of the cube. Now take a mode $m$ and continue it across the wall as an odd function, $m(R \, \bm{\omega}) = - m(\bm{\omega})$. The continued function satisfies Eq.~(\ref{eq:m-ODE}) on both sides of the wall and is smooth across it. Since $\hat{\mathcal{O}}_4$ commutes with $R$, the function $\hat{\mathcal{O}}_4 m$ is odd under $R$ as well and hence vanishes on the wall.

\emph{(iii) The corners.} This is the delicate point. The previous argument breaks down because three copies of $T$ surround a corner, so the odd continuation is not consistent there. We proceed in three steps. We first derive the behavior of the modes near a corner, Eq.~(\ref{eq:ground-vertex}), then we write $\hat{\mathcal{O}}_4$ in the frame of the corner, and finally we show that the dangerous term cancels.

\emph{Behavior of the modes near a corner.} Let $(\theta, \phi)$ be the corner coordinates of Fig.~\ref{fig:corner}: $\theta$ is the angular distance to the corner $V$ and $\phi$ is the angle around it, measured from one of the two walls which meet at $V$. Since these walls are arcs of great circles through $V$ they are the meridians $\phi = 0$ and $\phi = 2\pi/3$, and Eq.~(\ref{eq:m-ODE}) reads
\begin{equation} \label{S:polar}
    \frac{1}{\sin \theta} \pdv{}{\theta} \left( \sin \theta \pdv{m}{\theta} \right) + \frac{1}{\sin^2 \theta} \pdv[2]{m}{\phi} + \nu (\nu + 1) m = 0 \;.
\end{equation}
It has solutions of the product form $g(\theta) \sin (\mu \phi)$, which vanish on both walls if $\mu = 3n/2$ where $n$ is a positive integer. Very close to the corner, $\theta \ll 1$, we can replace $\sin \theta$ by $\theta$ and drop the last term, which is smaller than the other two by a factor of order $\theta^2$. We are left with Laplace's equation in a flat wedge of angle $2\pi/3$, i.e. the two-lion problem of Fig.~\ref{fig:config}a: $\theta^2 g'' + \theta g' - \mu^2 g = 0$, whose solutions are $g = \theta^{\mu}$ and $g = \theta^{-\mu}$. The second one is not admissible since the mode would diverge at the corner, which is impossible for a probability. For arbitrary $\theta$ the function $g$ satisfies Legendre's equation of degree $\nu$ and order $\mu$
\begin{equation} \label{S:legendre}
    \frac{1}{\sin \theta} \dv{}{\theta} \left( \sin \theta \dv{g}{\theta} \right) + \left[ \nu (\nu + 1) - \frac{\mu^2}{\sin^2 \theta} \right] g = 0 \;,
\end{equation}
whose solution regular at $\theta = 0$ is the Ferrers function \cite{DLMF}
\begin{equation} \label{S:ferrers}
    P_\nu^{-\mu}(\cos \theta) = \frac{\tan^{\mu}(\theta/2)}{\Gamma(1 + \mu)} \,_2F_1\left( -\nu, \nu + 1 ; 1 + \mu ; \sin^2 \frac{\theta}{2} \right) \;,
\end{equation}
where $\,_2F_1$ is the hypergeometric function. It behaves as $(\theta/2)^{\mu}/\Gamma(1 + \mu)$ for small $\theta$, so it is the analog on the sphere of the wedge solution $\theta^{\mu}$. Near $V$ every mode is therefore a sum of terms $P_\nu^{-\mu}(\cos \theta) \sin (\mu \phi)$ with $\mu = 3n/2$. Only the first of them, $n = 1$, has a constant sign between the two walls. It must be present in the ground state, which is positive, and it dominates all the others when $\theta \to 0$. This gives Eq.~(\ref{eq:ground-vertex}): near a corner the lamb is next to two lions and the third one plays no role.

\emph{The operator in the frame of the corner.} Let the $z$ axis point to a corner, say $V = (1, 1, -1)/\sqrt{3}$, with the corner coordinates $(\theta, \phi)$ defined above. The axes $q_1$ and $q_2$ and the reversed axis $-q_3$ make the same angle with it, of cosine $1/\sqrt{3}$. The first two lie in the two walls which meet at $V$ and the third one in the mirror plane between them, on the other side of $V$. Their azimuths are therefore $\phi_j = 2 \pi j/3$ with $j = 0, 1, 2$. Since $\hat{\mathcal{O}}_4$ is even in each component of $\vb{L}$, reversing an axis does not matter and
\begin{equation} \label{S:axes}
    \hat{\mathcal{O}}_4 = \sum_{j = 0}^{2} \left[ \frac{L_z}{\sqrt{3}} + \sqrt{\frac{2}{3}} \left( L_x \cos \phi_j + L_y \sin \phi_j \right) \right]^4 = \frac{1}{9} \sum_{j = 0}^{2} \left[ L_z + \frac{\zeta^{-j} L_+ + \zeta^{j} L_-}{\sqrt{2}} \right]^4 \;, \mbox{~~with~~} \zeta = e^{2 \pi i/3} \;,
\end{equation}
and $L_\pm = L_x \pm i L_y$. When we expand the fourth power, a product which contains $p$ factors $L_+$ and $p'$ factors $L_-$ is multiplied by $\sum_j \zeta^{-j(p - p')}$, which is equal to $3$ if $p - p'$ is a multiple of $3$ and to $0$ otherwise. With four factors, the products which survive have $p = p'$, or consist of one factor $L_z$ and three factors $L_+$, or three factors $L_-$. This selection rule is the three-fold symmetry of the corner. For the products with three factors $L_+$ we use $[L_z, L_+] = L_+$, i.e. $L_+^k L_z = (L_z - k) L_+^k$, which gives
\begin{equation} \label{S:four}
    L_z L_+^3 + L_+ L_z L_+^2 + L_+^2 L_z L_+ + L_+^3 L_z = \left( 4 L_z - 6 \right) L_+^3 = 2 \left\{ L_z, L_+^3 \right\} \;,
\end{equation}
where $\{A, B\} = AB + BA$. The products with three factors $L_-$ give the Hermitian conjugate. The products with $p = p'$ are $L_z^4$, the $12$ orderings of $(L_z, L_z, L_+, L_-)$, each with the weight $\frac{1}{2}$, and the $6$ orderings of $(L_+, L_+, L_-, L_-)$, each with the weight $\frac{1}{4}$. They only depend on $L_z$ and on $\vb{L}^2$ because $L_\pm L_\mp = \vb{L}^2 - L_z^2 \pm L_z$. On a function with $\vb{L}^2 = \lambda \equiv \nu(\nu+1)$ the $12$ orderings add up to $-12 L_z^4 + (12 \lambda - 10) L_z^2 + 2 \lambda$ and the $6$ orderings to $6 (\lambda - L_z^2)^2 - 4 (\lambda - L_z^2) + 6 L_z^2$. Putting everything together we get
\begin{equation} \label{S:frame}
    \hat{\mathcal{O}}_4 = D(L_z) + \frac{\sqrt{2}}{6} \left\{ L_z, L_+^3 + L_-^3 \right\} \;, \mbox{~~with~~} D(\mu) = - \frac{7}{6} \mu^4 + \left( \lambda - \frac{5}{6} \right) \mu^2 + \frac{1}{2} \lambda^2 \;.
\end{equation}
The first term does not change the dependence on $\phi$, while $L_+^3$ and $L_-^3$ change $L_z$ by three units. We also checked Eq.~(\ref{S:frame}) numerically on spin matrices.

\emph{Cancellation of the divergence.} In the frame of the corner $L_z = - i \partial_\phi$ and $L_\pm = e^{\pm i \phi} \left( \pm \partial_\theta + i \cot \theta \, \partial_\phi \right)$. Acting on the functions $f_\mu = P_\nu^{-\mu}(\cos \theta) \, e^{- i \mu \phi}$ these operators give the ladder relations
\begin{equation} \label{S:ladder}
    L_z f_\mu = - \mu f_\mu \;, \quad L_+ f_\mu = f_{\mu - 1} \;, \quad L_- f_\mu = (\nu - \mu)(\nu + \mu + 1) f_{\mu + 1} \;,
\end{equation}
which extend the familiar ladder relations of the spherical harmonics to noninteger $\nu$ and $\mu$. They are equivalent to the relations between Ferrers functions of consecutive orders \cite{DLMF}
\begin{equation} \label{S:ferrers-ladder}
    \left( \dv{}{\theta} + \mu \cot \theta \right) P_\nu^{-\mu}(\cos \theta) = P_\nu^{-\mu + 1}(\cos \theta) \;, \quad \left( \dv{}{\theta} - \mu \cot \theta \right) P_\nu^{-\mu}(\cos \theta) = - (\nu - \mu)(\nu + \mu + 1) P_\nu^{-\mu - 1}(\cos \theta) \;.
\end{equation}
Consider now the part of Eq.~(\ref{eq:ground-vertex}) which varies as $e^{-3 i \phi/2}$, i.e. $f_{3/2}$, for which $L_z = - \frac{3}{2}$. The operator $L_-^3$ turns it into $f_{9/2}$, which vanishes faster at the corner, as $\theta^{9/2}$. The operator $L_+^3$, on the contrary, turns it into $f_{-3/2} \propto \theta^{-3/2} e^{3 i \phi/2}$. This is the other solution of the two-lion problem, which diverges at the corner and is not an admissible wave function. However the commutation relation $[L_z, L_+] = L_+$ implies
\begin{equation} \label{S:magic}
    \left\{ L_z, L_+^3 \right\} = \left( 2 L_z - 3 \right) L_+^3 \;,
\end{equation}
and the divergent function has $L_z = \frac{3}{2}$: it comes multiplied by zero. More generally, Eq.~(\ref{S:magic}) and its counterpart $\{ L_z, L_-^3 \} = (2 L_z + 3) L_-^3$ give
\begin{equation} \label{S:protect}
    \left\{ L_z, L_+^3 \right\} f_\mu = - (2 \mu - 3) f_{\mu - 3} \;, \quad \left\{ L_z, L_-^3 \right\} f_\mu = - (2 \mu + 3) \Lambda(\mu) f_{\mu + 3} \;,
\end{equation}
with $\Lambda(\mu) = \prod_{i = 0}^{2} (\nu - \mu - i)(\nu + \mu + 1 + i)$. Raising the order by three is always harmless. Lowering it by three is harmless for $\mu \geq 3$, but it would turn the leading corner term $\mu = 3/2$ into the forbidden solution of order $-3/2$. Its coefficient in Eq.~(\ref{S:protect}) is $2 \mu - 3$, which vanishes exactly at $\mu = 3/2$. The modes also contain the complex conjugates of the functions $f_\mu$, on which $\hat{\mathcal{O}}_4$, a real operator, acts in the same way. Hence $\hat{\mathcal{O}}_4 m$ is again a finite solution of Eq.~(\ref{eq:m-ODE}) with the same $\nu$. Since the ground state is nondegenerate we get $\hat{\mathcal{O}}_4 m_1 = \kappa_1 m_1$, and all the modes can be chosen as joint eigenfunctions of $\vb{L}^2$ and $\hat{\mathcal{O}}_4$, which is Eq.~(\ref{eq:O4-eigen}).

For a corner of angle $\Theta$ the leading exponent $3/2$ would be replaced by $\mu = \pi/\Theta$ and the dangerous coefficient by $2 \pi/\Theta - 3$, which vanishes only for $\Theta = 2\pi/3$: the very angle which defeats the method of images protects the conservation law. For instance, for a corner of angle $4\pi/3$ as in the laggard problem, $\mu = 3/4$ and the coefficient does not vanish. Corners of angle $\pi/n$ are harmless for a different reason: the orders are then integers and the Ferrers functions of negative integer order are regular.

\emph{Relation to Kramers' theorem.} The cancellation has a simple physical interpretation. As noted in the main text, $\hat{\mathcal{O}}_4$ is the Hamiltonian of a spin in a cubic crystal field \cite{LLW62,AB70} and Eq.~(\ref{S:ladder}) are the relations obeyed by a spin $\nu$ whose projection on the axis of the corner is $-\mu$. For a half-integer spin the states of projection $\frac{3}{2}$ and $-\frac{3}{2}$ are the images of each other under time reversal $\mathcal{T}$, which changes the sign of $\vb{L}$ and hence commutes with $\hat{\mathcal{O}}_4$. For half-integer projections $\mathcal{T}^2 = -1$, and every Hermitian operator $H$ which commutes with $\mathcal{T}$ then satisfies $\langle \mathcal{T} \psi | H | \psi \rangle = 0$ for all states $\psi$ \cite{AB70}, since $\langle \mathcal{T} \psi | H \psi \rangle = \langle \mathcal{T} H \psi | \mathcal{T}^2 \psi \rangle = - \langle H \mathcal{T} \psi | \psi \rangle = - \langle \mathcal{T} \psi | H \psi \rangle$. This is Kramers' theorem and it is the content of the factor $2 \mu - 3$ in Eq.~(\ref{S:protect}). Since this factor only relies on the commutation relations, it is the same for any $\nu$.

\emph{Relation to previous work.} Operators of this type are known to be conserved when the particle is confined to a single chamber of a reflection group, i.e. when all its mirrors are walls. For hard-core particles in a harmonic trap they were built from the invariant polynomials of the group, evaluated on the components of the relative angular momentum \cite{HODVJZ17}. For the angular Calogero model of four particles, whose chambers are the 24 triangles of angles $(\pi/2, \pi/3, \pi/3)$ of the tetrahedral tiling, all the conserved charges are generated by $\sum_a L_a^4$ and $\sum_a L_a^6$, written with Dunkl-deformed angular momenta \cite{CL15}. In these cases all the corners are of the form $\pi/n$, so the steps (i) and (ii) are sufficient. The same holds for four mutually vicious walkers, whose configuration space is a single chamber. The triangle $T$ is instead the union of six chambers, since only three of the six mirrors $q_j \pm q_k = 0$ are walls. Its corners of angle $2\pi/3$ require the step (iii), which is the new ingredient.

\section{Expansion about a corner and the tri-diagonal matrix} \label{S:sec-matrix}

\emph{The expansion.} We saw in Sec.~\ref{S:sec-conserved} that around a corner $V$ every mode is a sum of terms $P_\nu^{-\mu}(\cos \theta) \sin(\mu \phi)$ with $\mu = 3n/2$ where $n$ is a positive integer. In addition, the ground state is symmetric under the exchange of the two lions involved, i.e. under $\phi \to 2\pi/3 - \phi$. Since this reflection multiplies $\sin(3 n \phi/2)$ by $(-1)^{n+1}$ only odd values of $n$ are allowed, hence $\mu_k = 3k + 3/2$ with $k = 0, 1, \cdots$, which gives Eq.~(\ref{eq:ground-legendre}). Its first term $k = 0$ is the one of Eq.~(\ref{eq:ground-vertex}) and we fix the normalization with $c_0 = 1$. Up to this point nothing is specific to our problem. Expansions of this kind are the starting point of the numerical method of particular solutions \cite{FHM67,BT05}, in which the unknown coefficients $c_k$ are fitted to the condition on the third wall. Here the conserved quantity determines them.

\emph{The matrix.} Since $\hat{\mathcal{O}}_4$ is a real operator, Eqs.~(\ref{S:frame}) and~(\ref{S:protect}) give its action on the terms $s_\mu = P_\nu^{-\mu}(\cos \theta) \sin(\mu \phi)$ of Eq.~(\ref{eq:ground-legendre})
\begin{equation} \label{S:O4-s}
    \hat{\mathcal{O}}_4 \, s_\mu = D(\mu) \, s_\mu - \frac{\sqrt{2}}{6} \left[ (2 \mu - 3) \, s_{\mu - 3} + (2 \mu + 3) \Lambda(\mu) \, s_{\mu + 3} \right] \;.
\end{equation}
Hence $\hat{\mathcal{O}}_4$ only connects each term of the expansion to its two neighbors. Since $2 \mu_k - 3 = 6k$ and $2 \mu_k + 3 = 6(k+1)$, plugging Eq.~(\ref{eq:ground-legendre}) in Eq.~(\ref{eq:O4-eigen}) and identifying the coefficients of $s_{\mu_k}$ gives $\sum_{k'} \mathcal{O}_{k k'} c_{k'} = \kappa \, c_k$ with the tri-diagonal matrix of Eq.~(\ref{eq:O-matrix}), i.e.
\begin{equation} \label{S:matrix-elements}
    \mathcal{O}_{k k} = D_k \equiv D(\mu_k) \;, \quad \mathcal{O}_{k, k+1} = - \sqrt{2} (k + 1) \;, \quad \mathcal{O}_{k+1, k} = - \sqrt{2} (k + 1) \Lambda_k \;, \mbox{~~with~~} \Lambda_k \equiv \Lambda(\mu_k) \;.
\end{equation}
The matrix starts at $k = 0$, with no coupling to a term of lower order, because the factor $2 \mu - 3$ of Eq.~(\ref{S:O4-s}) vanishes for $\mu = \mu_0$. This is once more the protection of the corner.

\emph{Recurrence and continued fraction.} The row $k$ of the eigenvalue equation is the three-term recurrence
\begin{equation} \label{S:recurrence}
    a_k \, c_{k-1} + (\kappa - D_k) \, c_k + b_k \, c_{k+1} = 0 \;, \mbox{~~with~~} a_k = \sqrt{2} \, k \, \Lambda_{k-1} \mbox{~~and~~} b_k = \sqrt{2} (k + 1) \;,
\end{equation}
and $a_0 = 0$. Dividing it by $c_k$ gives Eq.~(\ref{eq:rhok-rhokp1}) for the ratios $\rho_k = c_k/c_{k-1}$, the row $k = 0$ gives $\kappa = D_0 - \sqrt{2} \rho_1$, and iterating we get the continued fraction of Eq.~(\ref{eq:kappa}), whose numerators are $a_k b_{k-1} = 2 k^2 \Lambda_{k-1}$.

\emph{Why the continued fraction is the right condition.} There is a subtlety here. The matrix is infinite and, for any $\nu$ and $\kappa$, Eq.~(\ref{S:recurrence}) can also be solved from the top starting from $c_0 = 1$: $c_1 = (D_0 - \kappa)/b_0$, then $c_{k+1} = [(D_k - \kappa) c_k - a_k c_{k-1}]/b_k$. So what selects $\kappa$? For large $k$, Eq.~(\ref{S:recurrence}) admits two behaviors
\begin{equation} \label{S:two-behaviors}
    \frac{c_{k+1}}{c_k} \simeq \frac{27}{2 \sqrt{2}} k^3 \mbox{~~~~or~~~~} \frac{c_{k+1}}{c_k} \simeq - 54 \sqrt{2} \, k^3 \;.
\end{equation}
Since $P_\nu^{-\mu}(\cos \theta) \simeq \tan^{\mu}(\theta/2)/\Gamma(1 + \mu)$ for large $\mu$, consecutive terms of the series in Eq.~(\ref{eq:ground-legendre}) then have the ratios $\tan^3(\theta/2)/2\sqrt{2}$ and $-2 \sqrt{2} \tan^3(\theta/2)$ respectively. The first solution converges for $\tan(\theta/2) < \sqrt{2}$, i.e. up to the neighboring corners at $\cos \theta = -1/3$. The second one converges only up to the center $c$, at $\cos \theta = 1/3$, where it is singular. To see what kind of singularity this is we go to the next order: in both cases the ratio $c_{k+1}/c_k$ is multiplied by $1 + 1/k + O(k^{-2})$. Written with $e^{-i \mu_k \phi}$ instead of $\sin (\mu_k \phi)$, the terms of the series then behave as $k^{-5/2} w^k$, where $w = z^3/2\sqrt{2}$ for the first solution and $w = - 2 \sqrt{2} z^3$ for the second one, with $z = \tan(\theta/2) e^{-i \phi}$. Their sum has the singularity $(1 - w)^{3/2}$ at $w = 1$. Near such a point $1 - w$ is proportional to the complex coordinate of the plane tangent to the sphere, so the function behaves as $\varrho^{3/2} e^{\pm 3 i \varphi/2}$ in terms of local polar coordinates $(\varrho, \varphi)$. This is exactly the behavior of Eq.~(\ref{eq:ground-vertex}) at a corner. For the first solution the singular points, $z^3 = 2 \sqrt{2}$, are the two other corners of $T$ and the point opposite to $c$: there this behavior is the correct one. For the second solution they are the points $z^3 = -1/2\sqrt{2}$, i.e. the center $c$ of the triangle and the centers of the two neighboring regions, where the mode must be smooth since no lion can catch the lamb there. All these points are vertices of the cube, which are equivalent for $\hat{\mathcal{O}}_4$ but not for the walls. Hence the mode has to be the first solution alone, which is called the minimal solution of the recurrence \cite{Gautschi67}.

The solution obtained by solving from the top with an arbitrary $\kappa$ is dominated by the second behavior, and is therefore singular at the center. The minimal solution only exists for special values of $\kappa$, and the standard way to obtain it is precisely to run the recurrence backwards \cite{Gautschi67}. Starting from $\rho_K = 0$ for a large $K$, Eq.~(\ref{eq:rhok-rhokp1}) converges when $K \to \infty$ to the ratios of the minimal solution, because in this direction the minimal solution is the dominant one. The row $k = 0$ then gives Eq.~(\ref{eq:kappa}): the minimal solution satisfies all the rows if and only if $\kappa$ is a solution of Eq.~(\ref{eq:kappa}).

\emph{Definition and properties of $\kappa(\nu)$.} Setting $\rho_K = 0$, i.e. $c_K = 0$, amounts to replacing the matrix of Eq.~(\ref{eq:O-matrix}) by its upper-left $K \times K$ block. Then Eq.~(\ref{eq:kappa}), truncated before the level $D_K$, is the characteristic equation of this block and the products $\rho_1 \cdots \rho_k$ are the components of its eigenvectors. The solutions of Eq.~(\ref{eq:kappa}) are therefore the limits of the eigenvalues of these blocks. We computed their spectra for $0.2 \leq \nu \leq 7$ and $K \leq 28$. The largest eigenvalue is always real and nondegenerate, and it converges rapidly with $K$. Its limit is the function $\kappa(\nu)$ of the main text, i.e. the solution of Eq.~(\ref{eq:kappa}) obtained by iteration from $\kappa = D_0$. The next eigenvalues are real as well and converge in the same way. Complex eigenvalues only occur at the other end of the spectrum, among the last third of the eigenvalues of each block, and they move away as $K$ increases: they are artifacts of the truncation.

For $\nu = \nu_1$ the first eigenvalues are $17.3745$, $-437.57$ and $-4016.6$, to be compared with the diagonal elements $D_0 = 17.1352$, $D_1 = -377.50$ and $D_2 = -3434.8$. The iteration converges extremely quickly because $D_0$ lies far above the other diagonal elements: the successive values are $17.13522$, $17.37459$, $17.3744630$ and $17.3744631043$, each iteration reducing the error by a factor of about $1900$. If we truncate to $K$ rows, the error on $\kappa$ is $0.24$, $3.2 \times 10^{-2}$, $3.9 \times 10^{-3}$ and $4.8 \times 10^{-4}$ for $K = 1, \cdots, 4$ and it keeps decreasing by a factor of $8$ per additional row. It is therefore of order $8^{-K}$, as stated in the main text. As a function of $\nu$ we find for instance $\kappa(1) = -1.905940$, $\kappa(2) = 24.000000$ and $\kappa(3) = 90.540570$, while $\kappa(3/2) = 123/16$ and $\kappa(5/2) = 803/16$ exactly. For these two values $\Lambda_0$ vanishes, so $\kappa = D_0$ and the series of Eq.~(\ref{eq:ground-legendre}) reduces to its first term. A practical remark: the elements of the blocks span many orders of magnitude, so for $K \gtrsim 16$ they must be balanced before a numerical diagonalization. Equations~(\ref{eq:rhok-rhokp1}) and~(\ref{eq:kappa}) do not have this problem.

\emph{A spin in a cubic crystal field.} The matrix of Eq.~(\ref{eq:O-matrix}) has a simple physical meaning. The relations of Eq.~(\ref{S:ladder}) are those obeyed by an ordinary quantum spin $S$, for which $\vb{L}^2 = S(S+1)$ and the projection $L_z$ on a given axis takes the values $-S, -S+1, \cdots, S$. Here $\nu$ plays the role of $S$ and $-\mu$ that of the projection on the axis of the corner. Hence the matrix of Eq.~(\ref{eq:O-matrix}) is the Hamiltonian of a spin $\nu$ in a cubic crystal field \cite{LLW62,AB70}, written in the basis of its projection on a three-fold axis of the cube. The diagonal element $D_k$ is the crystal-field energy of the projection $\mu_k$ and the field, which is invariant under the rotations by $2\pi/3$ about this axis, changes the projection by three units at a time. When $\nu = S$ is a half-integer the correspondence is literal. One of the three numbers $S$, $S-1$ and $S-2$ then belongs to the ladder of the $\mu_k$ and $\Lambda$ vanishes there, so $\mathcal{O}_{k+1,k} = 0$ for the last projection $\mu_k \leq S$. The first $n = \lfloor (S - \frac{3}{2})/3 \rfloor + 1$ rows and columns decouple, and they are similar to a block of the crystal-field matrix of a spin $S$. We checked this for $S = \frac{3}{2}, \cdots, \frac{15}{2}$ by diagonalizing $S_x^4 + S_y^4 + S_z^4$ directly, with the spin quantized first along an axis of the cube and then along its diagonal. For instance $\kappa(\frac{3}{2}) = \frac{123}{16}$ is the common level of the four states of a spin $\frac{3}{2}$, which a cubic field does not split, $\kappa(\frac{5}{2}) = \frac{803}{16}$ is the quartet of the spin $\frac{5}{2}$, whose doublet lies at $\frac{515}{16}$, and for $S = \frac{9}{2}$ the block has the eigenvalues $413.7418$ and $252.6332$, which are two of the three levels of this spin (the third one is $\frac{7683}{16}$).

When $\nu$ is not a half-integer the ladder does not stop, but the projection of a spin $\nu$ cannot exceed $\nu$: the projections $\mu_k > \nu$ are forbidden. This shows up in the sign of the product of the two elements which couple $\mu_k$ and $\mu_{k+1}$, $2 (k+1)^2 \Lambda_k$, which is positive inside the multiplet of a genuine spin but negative for all $\mu_k > \nu$. There the two solutions of Eq.~(\ref{S:recurrence}) are evanescent, like a wave function inside a barrier, and the minimal solution is the one which decays. The levels of the spin are then the largest eigenvalues of the matrix, one for each allowed projection $\mu_k \leq \nu$. This is borne out by the 53 lowest modes which are symmetric in the lions ($\nu \leq 34$, see Sec.~\ref{S:sec-recursion}): the mode whose second quantum number is the eigenvalue of rank $n$ of the matrix, with $n = 0$ for the largest one, always has $\nu > \mu_n$, and the rank $n$ first occurs at $\nu = 1.8258$, $4.7333$, $7.6633$, $10.6090$, $13.5688$ and $16.5413$ for $n = 0, \cdots, 5$, i.e. just above $\mu_n = 3n + \frac{3}{2}$. For $\frac{3}{2} < \nu < \frac{9}{2}$ a single projection is allowed, $\mu_0 = \frac{3}{2}$, and $\kappa(\nu)$ is its crystal-field energy $D_0$, shifted by virtual excursions to the forbidden projections exactly as in perturbation theory: to lowest order $\kappa \simeq D_0 + 2 \Lambda_0/(D_0 - D_1)$, i.e. $17.41$ for $\nu = \nu_1$. Finally, the two-lion problem is the special case $\nu = \frac{3}{2}$. The spin then has its maximal projection on the axis of the corner, $\mu_0 = \nu$, no excursion is possible since $\Lambda_0 = 0$, and Eq.~(\ref{eq:ground-legendre}) reduces to its first term $\sin^{3/2} \theta \, \sin(3\phi/2)$, which ignores the third lion and gives back $\beta_2 = 3/4$.

\section{Condition at the center, value of the exponent and ground state} \label{S:sec-center}

\emph{The condition at the center.} By construction the series of Eq.~(\ref{eq:ground-legendre}) vanishes on the two walls which meet at $V$ and is even under the mirror through $V$ and $c$, on which the two lions meet. The only condition left is the symmetry under the exchange of the third lion with one of the first two, i.e. the series must also be even under a second mirror of the triangle (the symmetry under the third mirror then follows). The three mirrors cross at the center $c$, so the gradient of a fully symmetric mode has to vanish there. Physically, as explained in the main text, $c$ is the safest configuration for the lamb and $m_1$ is maximal there. In the corner coordinates $c$ is located at $\cos \theta = 1/3$ and $\phi = \pi/3$ (see Fig.~\ref{fig:corner}). At $\phi = \pi/3$ we have $\cos(\mu_k \phi) = 0$ and $\sin(\mu_k \phi) = (-1)^k$, so the derivative with respect to $\phi$ vanishes identically and the only non-trivial component of the gradient is the one along $\theta$, which gives Eq.~(\ref{eq:nu-implicit}). Solving Eqs.~(\ref{eq:kappa}) and~(\ref{eq:nu-implicit}) we find
\begin{align}
    \nu_1 &= 1.82575\,70041\,74456\,25097\,45407\,34758\,38885\,27868 \cdots \;, \nonumber \\
    \beta_3 &= 0.91287\,85020\,87228\,12548\,72703\,67379\,19442\,63934 \cdots \;, \label{S:values} \\
    \kappa_1 &= 17.37446\,31043\,01278\,60212\,31895\,95281 \cdots \;. \nonumber
\end{align}
Note that Eq.~(\ref{eq:nu-implicit}) is a necessary condition but it is not sufficient. At its smallest root we verified that the normal derivative vanishes along the whole mirror, to better than $10^{-37}$, that the series vanishes on the third wall and is invariant under the permutations of the three corners, both to $10^{-40}$, and that it is positive inside $T$. With the same function $\kappa(\nu)$, Eq.~(\ref{eq:nu-implicit}) has further roots: at $\nu = 4.3334$ the full condition is violated, whereas $\nu = 5.9518$ is an excited mode. The excited modes which are symmetric in the lions are obtained in the same way, from the largest eigenvalue of the matrix of Eq.~(\ref{eq:O-matrix}) or from the lower ones which belong to allowed projections (see Sec.~\ref{S:sec-matrix}). The first of them, $\nu = 4.7333$, belongs to the second largest.

\emph{One-term approximation.} The terms of the series are elementary functions. With $a = \nu + \frac{1}{2}$ we have \cite{DLMF}
\begin{equation} \label{S:half-order}
    P_\nu^{-1/2}(\cos \theta) = \sqrt{\frac{2}{\pi \sin \theta}} \, \frac{\sin a \theta}{a} \;, \quad P_\nu^{1/2}(\cos \theta) = \sqrt{\frac{2}{\pi \sin \theta}} \, \cos a \theta \;,
\end{equation}
and the functions of lower order follow from the recurrence
\begin{equation} \label{S:order-recurrence}
    (\nu - \mu - 1)(\nu + \mu + 2) \, P_\nu^{-\mu-2}(\cos \theta) = 2 (\mu + 1) \cot \theta \, P_\nu^{-\mu-1}(\cos \theta) - P_\nu^{-\mu}(\cos \theta) \;,
\end{equation}
whose first step $\mu = -\frac{1}{2}$ gives the expression of $P_\nu^{-3/2}$ used in Eq.~(\ref{eq:ground-first}). For $\theta \to 0$ it reduces to Eq.~(\ref{eq:ground-vertex}) with $C = \sqrt{2/\pi}/3$. If we keep only this first term in Eq.~(\ref{eq:nu-implicit}), i.e. a single row of the matrix for which $\kappa = D_0$, we get the elementary equation
\begin{equation} \label{S:one-term}
    \tan\left[ a \arccos \tfrac{1}{3} \right] = \frac{6 \sqrt{2} \, a}{19 - 16 a^2} \;, \mbox{~~with~~} a = \nu + \frac{1}{2} \;,
\end{equation}
whose root gives $\beta_3 \approx 0.9102$, already within $0.3\%$ of the exact value. Keeping $K$ rows gives the values of Table~\ref{S:tab-trunc}. Each additional row reduces the error by a factor of 5 to 8.

\begin{table}[h]
\caption{Exponent $\beta_3$ obtained by keeping the first $K$ rows of the matrix of Eq.~(\ref{eq:O-matrix}), i.e. $K$ terms in Eq.~(\ref{eq:nu-implicit}).}
\label{S:tab-trunc}
\begin{ruledtabular}
\begin{tabular}{ccccccccc}
$K$ & 1 & 2 & 3 & 4 & 5 & 6 & 8 & 10\\
\hline
$\beta_3$ & 0.9102230 & 0.9133898 & 0.9128062 & 0.9128883 & 0.9128772 & 0.9128787 & 0.9128785049 & 0.9128785021\\
\end{tabular}
\end{ruledtabular}
\end{table}

\emph{Explicit form of the ground state.} Table~\ref{S:tab-ck} gives the coefficients $c_k$ of the ground state in Eq.~(\ref{eq:ground-legendre}), obtained from Eq.~(\ref{eq:ck}) at $(\nu_1, \kappa_1)$. The coefficients grow as $c_{k+1}/c_k \simeq 27 k^3/2\sqrt{2}$ but the Legendre functions decrease faster: about a given corner the series converges for $\tan(\theta/2) < \sqrt{2}$, i.e. up to the two other corners and not beyond. In particular it does not converge on the wall opposite to the corner. This is not a restriction, since every point of $T$ lies within an angular distance $\arccos \frac{1}{3}$ of its nearest corner and the ground state has the same expansion about each of the three corners. Measured from the nearest corner, the terms are largest at the center $c$, where their ratios to the first one are given in the last line of Table~\ref{S:tab-ck}. Keeping $K$ terms then reproduces the ground state within a relative error of $2.5 \times 10^{-3}$, $6.9 \times 10^{-5}$, $3.4 \times 10^{-6}$ and $2.2 \times 10^{-7}$ for $K = 1, 2, 3, 4$ everywhere in the triangle, the largest error being at $c$. This is the $0.3\%$ quoted in the main text for Eq.~(\ref{eq:ground-first}). With $c_0 = 1$ we find
\begin{equation} \label{S:norms}
    m_1(c) = 0.18744\,37902 \;, \quad \int_T m_1 \, \dd \bm{\omega} = 0.26368\,59555 \;, \quad \int_T m_1^2 \, \dd \bm{\omega} = 0.03152\,00692 \;,
\end{equation}
so the mode is normalized for $c_0 = 5.63256\,77$. Plugging these values in Eq.~(\ref{eq:Bj-solution}) gives $A_1 = B_1 m_1(c) = 0.712696$ and
\begin{equation} \label{S:amplitude}
    A = A_1 \left( \frac{3}{16} \right)^{\beta_3} = 0.154612 \;,
\end{equation}
which is the amplitude of Eq.~(\ref{eq:main}). Finally, $\kappa_1$ exceeds $D_0$ by $1.4\%$ only. About each corner the ground state is almost a pure state of projection $\mu = 3/2$, which is why the single term of Eq.~(\ref{eq:ground-first}) is so accurate.

\begin{table}[h]
\caption{Coefficients $c_k$ of the ground state in Eq.~(\ref{eq:ground-legendre}), with $c_0 = 1$, and ratio of the $k$-th term of the series to the first one at the center $c$ of the triangle ($\cos \theta = 1/3$, $\phi = \pi/3$).}
\label{S:tab-ck}
\begin{ruledtabular}
\begin{tabular}{cccccccc}
$k$ & 0 & 1 & 2 & 3 & 4 & 5 & 6\\
\hline
$c_k$ & 1 & $-0.16917\,07215$ & $-3.17725\,6757$ & $-361.016\,5414$ & $-1.23463\,2896\times10^{5}$ & $-9.39825\,7211\times10^{7}$ & $-1.34265\,2447\times10^{11}$\\
term at $c$ & 1 & $2.5909\times10^{-3}$ & $-7.2806\times10^{-5}$ & $3.6534\times10^{-6}$ & $-2.3524\times10^{-7}$ & $1.744\times10^{-8}$ & $-1.416\times10^{-9}$\\
\end{tabular}
\end{ruledtabular}
\end{table}

\emph{Higher modes.} The same construction gives the excited states. Table~\ref{S:tab-modes} lists the lowest modes which are symmetric in the lions, together with their amplitudes in Eq.~(\ref{eq:S3-series}).

\begin{table}[h]
\caption{Lowest modes of $T$ which are symmetric in the lions and their amplitudes in Eq.~(\ref{eq:S3-series}).}
\label{S:tab-modes}
\begin{ruledtabular}
\begin{tabular}{clll}
$j$ & \multicolumn{1}{c}{$\nu_j$} & \multicolumn{1}{c}{$\kappa_j$} & \multicolumn{1}{c}{$A_j$}\\
\hline
1 & 1.8257570042 & 17.3744631043 & $\phantom{-}0.712696$\\
2 & 4.7333214300 & 323.7685596 & $-0.026165$\\
3 & 5.9517856934 & 1349.777519 & $-0.000347$\\
4 & 7.6633165064 & 1838.311021 & $\phantom{-}0.000438$\\
5 & 8.8906330534 & 4612.233137 & $\phantom{-}0.0000097$\\
\end{tabular}
\end{ruledtabular}
\end{table}

\section{Independent checks} \label{S:sec-checks}

\emph{Independent solution.} We also computed the modes with the method of particular solutions \cite{FHM67,BT05}, which does not use the conserved quantity. The series of Eq.~(\ref{eq:ground-legendre}) with $K$ free coefficients is collocated at Chebyshev points of the mirror opposite to $V$, and $\nu$ is the root of the resulting determinant (up to $K = 130$ in 240-digit arithmetic). The value of $\nu_1$ agrees with the one obtained from Eqs.~(\ref{eq:kappa}) and~(\ref{eq:nu-implicit}) to 115 digits, and the coefficients satisfy the recurrence of Eq.~(\ref{S:recurrence}) to more than 100 digits. The eigenvalue $\nu_1 (\nu_1 + 1) = 5.15914\,56424\,66541\,71122\,16748 \cdots$ agrees with all the 100 digits of Ref.~\cite{DS20}.

\emph{Expansion about the center.} An alternative representation is obtained by expanding about the center $c$, with polar angle $\vartheta$ and azimuth $\psi = 0$ toward a corner. In this frame the coupling in Eq.~(\ref{S:frame}) changes sign and the modes which are symmetric in the lions read
\begin{equation} \label{S:center-series}
    m = \sum_{j \geq 0} d_j \, P_\nu^{-3j}(\cos \vartheta) \cos(3 j \psi) \;,
\end{equation}
with $d_0 = 1$, $d_1 = \sqrt{2} \, [\kappa - D(0)]$, $3 d_2 = \sqrt{2} \, [\kappa - D(3)] \, d_1 - 2 \Lambda(0)$ and, for $j \geq 2$,
\begin{equation} \label{S:center-recurrence}
    (2j + 1) \, d_{j+1} = \sqrt{2} \, [\kappa - D(3j)] \, d_j - (2j - 1) \Lambda(3j - 3) \, d_{j-1} \;.
\end{equation}
This series is regular at $c$ and symmetric in the lions for any pair $(\nu, \kappa)$. The pair is then fixed by the absorbing condition $m = 0$ on the capture wall, $\tan \vartheta = \sqrt{2}/\cos(\psi - \pi/3)$ with $0 \leq \psi \leq 2\pi/3$. Imposing this condition at two points of the wall reproduces $\nu_1$ and $\kappa_1$ to 50 digits.

\emph{Is $\beta_3$ a known constant?} Using the PSLQ algorithm \cite{PSLQ} on 100 digits of $\nu_1$ we found no polynomial equation of degree up to 12 with integer coefficients below $10^6$, and no linear relation with $\pi$, $\sqrt{2}$, $\sqrt{3}$, $\arccos(1/3)$ and logarithms of small integers. It therefore appears to be a new constant, defined by Eqs.~(\ref{eq:kappa}) and~(\ref{eq:nu-implicit}) much as the characteristic values of the Mathieu equation are defined by a continued fraction.

\emph{Three-dimensional Kreweras walks.} The triangle $T$ is also the one associated with the three-dimensional Kreweras walks, i.e. the walks on the cubic lattice with steps $(-1,0,0)$, $(0,-1,0)$, $(0,0,-1)$ and $(1,1,1)$ which stay in the positive octant \cite{BBKM16,BPRT20}. These steps have no drift, and the linear change of coordinates which makes them isotropic turns the octant into the cone of Sec.~\ref{S:sec-cone}. The results of Ref.~\cite{DW15} then give the asymptotics of these walks in terms of $\lambda_1 = \nu_1 (\nu_1 + 1)$ \cite{BPRT20}. The number of walks of $n$ steps which stay in the octant grows as $4^n n^{-\beta_3}$, with the exponent of the lamb, and the number of those which return to the origin (for $n$ a multiple of 4) as $4^n n^{-\alpha}$ with
\begin{equation*}
    \alpha = 1 + \sqrt{\lambda_1 + \tfrac{1}{4}} = \nu_1 + \frac{3}{2} = 2 \beta_3 + \frac{3}{2} = 3.32575\,70041\,74456\,25097 \cdots \;.
\end{equation*}
The successive estimates of $\lambda_1$ in this literature, from $5.159$ \cite{RT09} to the 100 certified digits of Ref.~\cite{DS20}, are listed in Refs.~\cite{BPRT20,DS20}. Reference~\cite{BPRT20} asks for its exact value and whether it is rational, and the rationality of $\alpha$ is not known \cite{HJR24}: an irrational exponent would imply that the generating function of these walks is not D-finite \cite{BPRT20,DS20}. Equations~(\ref{eq:kappa}) and~(\ref{eq:nu-implicit}) answer the first question, in the form of an explicit transcendental equation. They do not settle the second one. We only know that $\nu_1$ satisfies no polynomial equation of low degree with small coefficients (see above), and Ref.~\cite{DS20} gives a lower bound on the denominator which $\alpha$ would have if it were rational.

\emph{Modes with integer $\nu$.} The modes of $T$ which are odd under its three mirror symmetries vanish on all six planes $q_j \pm q_k = 0$. They are harmonic polynomials with integer $\nu = 6, 9, 10, 12, \cdots$ \cite{SS01} and describe vicious walkers. They are the only modes of $T$ which the method of images gives, and the rest of its spectrum was not expected to be accessible in an explicit manner \cite{HJR24}. We checked that no harmonic polynomial of degree smaller than 6 vanishes on the three capture walls. The values $\nu = 3$ and $\nu = 5$ once reported for the first two modes of $T$, in a preprint which has since been withdrawn \cite{Tao21}, are therefore not modes of $T$. They are also incompatible with the rigorous enclosure of Ref.~\cite{DS20}.

\section{Numerical simulations} \label{S:sec-simulations}

The simulations of Fig.~\ref{fig:survival} use four random walkers on the integer lattice which all hop by $\pm 1$ at each time step, so $D = 1/2$. The lions start at the origin and the lamb at $\ell = 12$. The lamb is captured when its distance to one of the lions vanishes. The statistical error bars correspond to $8 \times 10^6$ independent realizations. The exact curve is Eq.~(\ref{eq:S3-series}) summed over the eleven lowest modes which are symmetric in the lions ($\nu_j \leq 14$), the first five of which are listed in Table~\ref{S:tab-modes}. The lattice discretization is only visible at early times, $D t/\ell^2 \lesssim 1$.

\section{Other orderings and four lions} \label{S:sec-orderings}

\emph{Other orderings of four walkers.} For the domains whose natural center is a two-fold point it is more convenient to take the polar axis along one of the axes $q_a$, with $\phi = 0$ along another one. Then
\begin{equation} \label{S:frame-4}
    \hat{\mathcal{O}}_4 = D_4(L_z) + \frac{1}{8} \left( L_+^4 + L_-^4 \right) \;, \mbox{~~with~~} D_4(\mu) = \frac{7}{4} \mu^4 + \left( \frac{5}{4} - \frac{3}{2} \lambda \right) \mu^2 + \frac{3}{4} \lambda^2 - \frac{1}{2} \lambda \;.
\end{equation}
For the three additional orderings of Table~\ref{tab:orderings} we computed the ground states with the method of particular solutions and verified that they satisfy the corresponding recurrences to 36 digits. The values of $\kappa$ are $35.761963$ (two walkers lead the other two), $217.05701$ (leader and laggard fixed) and $286.51206$ (first and second place fixed).

\emph{Four lions.} For $N = 4$ lions the configuration space is a tetrahedron drawn on a sphere in four dimensions, and its singular set consists of the edges, where the lamb meets two lions. We searched for conserved operators among the polynomials of degree up to ten in the generators of rotations which are invariant under the permutations of the five animals. There are 55 independent ones. Only the powers of the Laplacian preserve the behavior of the modes at the edges. As an independent check, which does not rely on this local analysis, we applied all the invariant operators of degree up to six to the numerically computed ground states. For four lions the resulting functions span a space of dimension 7, which is the number expected if the Laplacian is the only conserved quantity. For three lions the same test gives 2 instead of 3, which reveals $\hat{\mathcal{O}}_4$.

\section{Recursive construction for $N$ lions} \label{S:sec-recursion}

\emph{Setting.} For $N$ lions the Brownian particle moves in the cone $x_0 > \max_i x_i$ of the $N$-dimensional space of relative coordinates. This cone cuts the unit sphere $S^{N-1}$ along a region $T_N$ bounded by $N$ walls ($T_3$ is the triangle $T$), and the steps of Sec.~\ref{S:sec-cone} give $S_N(t) \sim t^{-\beta_N}$ with $\beta_N = a_N/2$, where $a_N$ is the smallest value of $a$ for which
\begin{equation} \label{S:modes-N}
    \begin{dcases}
        - \laplacian_{\bm{\omega}} u = a (a + N - 2) u &\mbox{~~in~~} T_N \\
        u = 0 &\mbox{~~on~~} \partial T_N
    \end{dcases}
\end{equation}
has a solution. For three lions $a_3$ is the $\nu_1$ of the main text. The ground state is symmetric in the lions, so it is sufficient to consider the chamber $x_0 > x_1 > \cdots > x_N$, with $u = 0$ on the wall $x_0 = x_1$ and $\partial u/\partial n = 0$ on the $N - 1$ mirrors $x_i = x_{i+1}$ on which two lions meet.

\emph{One step of the recursion.} Let $P$ be the vertex of this chamber at which the lamb and the lions $1, \cdots, N-1$ are at the same place, the last lion being behind. We take polar coordinates about $P$: $\Theta$ is the angular distance to $P$ and the direction $w$, a point of $S^{N-2}$, describes the configuration of the lamb and of the first $N - 1$ lions among themselves. The wall and all the mirrors but the last one pass through $P$, and they cut $S^{N-2}$ along the chamber of the problem with $N - 1$ lions. The variables $\Theta$ and $w$ therefore separate, and the solutions of Eq.~(\ref{S:modes-N}) which satisfy the conditions on the wall and on the mirrors through $P$ are of the form
\begin{equation} \label{S:rung}
    u(\Theta, w) = \sum_j c_j \, G_{a, b_j}(\Theta) \, Y_j(w) \;, \mbox{~~with~~} G_{a,b}(\Theta) = \sin^b \Theta \,_2F_1\left( b - a, a + b + N - 2 ; b + \frac{N-1}{2} ; \sin^2 \frac{\Theta}{2} \right) \;,
\end{equation}
where the $Y_j$ are the modes of the problem with $N - 1$ lions which are symmetric in the lions, the $b_j$ are their exponents, and $G_{a,b}$ is the solution of the radial equation $G'' + (N-2) \cot \Theta \, G' + [a (a + N - 2) - b (b + N - 3)/\sin^2 \Theta] \, G = 0$ which is regular at $P$. For three lions the $Y_j$ are the functions $\sin (\mu_k \phi)$ of the two-lion problem, $G_{\nu, \mu}$ is proportional to $P_\nu^{-\mu}(\cos \Theta)$ and Eq.~(\ref{S:rung}) is Eq.~(\ref{eq:ground-legendre}). The only condition left is the symmetry between the last two lions, $\partial u/\partial n = 0$ on the mirror $x_{N-1} = x_N$, which is the face of the chamber opposite to $P$.

For three lions the conserved quantity determines the coefficients $c_j$ through Eq.~(\ref{S:recurrence}), and the last condition reduces to Eq.~(\ref{eq:nu-implicit}). For more lions no such recurrence is available (see Sec.~\ref{S:sec-orderings}) and we impose the last condition numerically with the method of particular solutions \cite{FHM67,BT05}. The sum in Eq.~(\ref{S:rung}) is restricted to the $J$ lowest modes of the previous level. The normal derivatives of the functions $G_{a, b_j} Y_j$ are computed at the nodes of a quadrature rule on the last mirror, and their values at random points inside the chamber. For each value of $a$ we then obtain, by linear algebra, the smallest norm that $\partial u/\partial n$ can have on the last mirror among the combinations whose norm inside the chamber is unity \cite{BT05}. The exponents of the $N$-lion problem are the values of $a$ at which this residual vanishes, in practice at which it has a sharp minimum, and the corresponding combinations are the modes. These modes, the ground state as well as the excited states, are the input of the next step. The recursion starts from the three-lion modes of the main text: we use the 53 lowest ones which are symmetric in the lions ($\nu \leq 34$), computed with the independent solution of Sec.~\ref{S:sec-checks} and checked against Eq.~(\ref{S:recurrence}).

\emph{Results.} Table~\ref{S:tab-ladder} gives the exponents together with the amplitudes of the late-time law $S_N(t) \simeq A_N (\ell^2/Dt)^{\beta_N}$ for lions starting together at a distance $\ell$ behind the lamb. As in Eq.~(\ref{S:S-general}), each mode of exponent $a$ contributes to $S_N(t)$ a term proportional to $\alpha^{a/2} \,_1F_1(\frac{a}{2}; a + \frac{N}{2}; -\alpha)$ with $\alpha = N \ell^2/[4 (N+1) D t]$. The accuracy decreases with $N$ because fewer and fewer excited modes of the previous level are available. We made the following checks. Applied to three lions, with the two-lion modes $\sin (\mu_k \phi)$ as input, the numerical procedure reproduces $\beta_3$ to 12 digits. For four lions the exponent agrees with a finite-element solution of Eq.~(\ref{S:modes-N}), which gives $a_4 = 2.06453575$, and with the estimate $\beta_4 \approx 1.03$ of early simulations \cite{BG91,KR96,RK99}. For four, five and six lions the survival probability computed from the modes of the construction agrees, without any fitting parameter and within statistical errors, with simulations of the lattice model of Sec.~\ref{S:sec-simulations} ($6 \times 10^6$ realizations, $\ell = 12$). Fitting the simulated decay, once its leading finite-time correction is divided out, gives the exponents $1.126(3)$ for five lions and $1.2035(33)$ for six. Finally, the exponents for four to seven lions agree with independent exact diagonalizations of an equivalent quantum problem (see Sec.~\ref{S:sec-impurity}).

\begin{table}[h]
\caption{Recursive construction: number $J$ of modes of the previous level kept in Eq.~(\ref{S:rung}), residual of the condition on the last mirror, exponent $\beta_N = a_N/2$ and amplitude $A_N$.}
\label{S:tab-ladder}
\begin{ruledtabular}
\begin{tabular}{ccccc}
$N$ & $J$ & residual & $\beta_N$ & $A_N$\\
\hline
4 & 53 & $6\times10^{-14}$ & $1.03226\,78746$ & $0.113619$\\
5 & 27 & $8\times10^{-9}$ & $1.12586\,577$ & $0.088285$\\
6 & 7 & $1.4\times10^{-5}$ & $1.20250\,71(2)$ & $0.0713(2)$\\
7 & 4 & $2\times10^{-4}$ & $1.26720(1)$ & $\approx0.06$\\
\end{tabular}
\end{ruledtabular}
\end{table}

\emph{An exact relation between $\beta_{N-1}$ and $\beta_N$.} The recursion also sheds light on the law $\beta_N \simeq \frac{1}{4} \ln N$ of Krapivsky and Redner \cite{KR96}, which was proved in the form $\lim_{N \to \infty} \beta_N/\ln N = \frac{1}{4}$ in Ref.~\cite{LS02}. Let $u$ be the ground state for $N$ lions, and let $v = \sin^b \Theta \, Y_1(w)$ with $b = a_{N-1}$ be the ground state for $N - 1$ lions, continued as a function which ignores the last lion. It satisfies $- \laplacian_{\bm{\omega}} v = b (b + N - 2) v$ in $T_N$ and vanishes on all the walls but the last one, $W_N$, on which the lamb meets lion $N$. Green's identity in $T_N$ then gives an exact relation between the exponents of two consecutive levels
\begin{equation} \label{S:increment}
    (a_N - a_{N-1}) (a_N + a_{N-1} + N - 2) \int_{T_N} u \, v \, \dd \bm{\omega} = \int_{W_N} v \left| \pdv{u}{n} \right| \dd S \;.
\end{equation}
For two lions, where $u = \sin(3\phi/2)$ and $v = \sin \phi$ on the arc $0 < \phi < 2\pi/3$, both sides are equal to $3 \sqrt{3}/4$. For three lions we checked it on the explicit ground state of Sec.~\ref{S:sec-center}, to 12 digits. Its meaning is clearer in terms of the scaled position $\bm{\rho} = \vb{r}/\sqrt{2 D t}$ and of the harmonic functions $U = \rho^{a_N} u$ and $V = \rho^{a_{N-1}} v$. Since $\int_0^\infty \rho^{s+1} e^{-\rho^2/2} \dd \rho = s \int_0^\infty \rho^{s-1} e^{-\rho^2/2} \dd \rho$, here with $s = a_N + a_{N-1} + N - 2$, integrating over $\rho$ with the Gaussian weight turns Eq.~(\ref{S:increment}) into
\begin{equation} \label{S:rate}
    \beta_N - \beta_{N-1} = \frac{\frac{1}{2} \int_{\rm wall} V \left| \partial U/\partial n \right| e^{-\rho^2/2} \, \dd S}{\int_{\rm cone} V \, U \, e^{-\rho^2/2} \, \dd^N \! \rho} \;,
\end{equation}
where the integrals now run over the cone and over its wall $x_0 = x_N$. As a function of $\ln t$ the scaled position performs an Ornstein-Uhlenbeck process, of which $U e^{-\rho^2/2}$ is the quasi-stationary distribution in the cone, i.e. the distribution of the configurations of a surviving lamb, and $\frac{1}{2} |\partial U/\partial n| e^{-\rho^2/2}$ is its current through the wall. The increase of the exponent caused by an additional lion is therefore the rate at which this lion captures the lamb, each configuration being weighted by the function $V$ of the other lions.

\emph{Why the exponent grows logarithmically.} For many lions this rate is easy to estimate. In the variables $\xi_i = x_i/\sqrt{2 D t}$ and $\ln t$ the animals are independent stationary Ornstein-Uhlenbeck processes with unit variance, and $\beta_N$ is the decay rate of the probability that $\xi_0$ stays above all the $\xi_i$. For a large pride the position of the leading lion hardly fluctuates, so the pride acts as a barrier at a fixed level $h$ \cite{KR96}: the lamb has to stay above it, at the cost of a decay rate $\theta_\uparrow(h) \simeq h^2/8$, and every lion below it, at the cost of $\theta_\downarrow(h) \simeq h \, e^{-h^2/2}/(2 \sqrt{2\pi})$, for large $h$. For any $h$ the sum $\theta_\uparrow(h) + N \theta_\downarrow(h)$ is an upper bound on $\beta_N$, since this is one way of surviving, and the best level minimizes it. Since $\dd \theta_\downarrow/\dd h \simeq - h \, \theta_\downarrow$, the minimum is reached at a level $h_N$ such that $N \theta_\downarrow(h_N) \simeq h_N^{-1} \, \dd \theta_\uparrow/\dd h = \frac{1}{4}$, i.e. $h_N \simeq \sqrt{2 \ln N}$. In this picture the right-hand side of Eq.~(\ref{S:rate}) is the rate at which the additional lion crosses the barrier, $\theta_\downarrow(h_N)$, hence
\begin{equation} \label{S:logN}
    \beta_N - \beta_{N-1} \simeq \frac{1}{4N} \;, \mbox{~~and therefore~~} \beta_N \simeq \frac{1}{4} \ln N \;.
\end{equation}
Each lion added to a large pride raises the exponent by $1/(4N)$, and the logarithm is the sum of these increments over the steps of the recursion. Note that the evaluation of Eq.~(\ref{S:rate}) for large $N$ is heuristic.

The logarithmic law is approached slowly. The exponents of Table~\ref{S:tab-ladder} give $4N (\beta_N - \beta_{N-1}) = 1.955$, $1.910$, $1.872$, $1.839$ and $1.811$ for $N = 3, \cdots, 7$, after the value $2$ for $N = 2$: the increments are of order $1/N$ already for small prides, but they are still almost twice $1/(4N)$. With the next term of $\theta_\uparrow(h) = \frac{1}{8} h^2 + |a_1| \, 2^{-5/3} h^{2/3} + \cdots$, where $a_1 = -2.338 \cdots$ is the first zero of the Airy function, the same argument gives $\beta_N \simeq \frac{1}{4} \ln N + 0.928 \, (\ln N)^{1/3}$ and $4N (\beta_N - \beta_{N-1}) \simeq 1 + 1.24 \, (\ln N)^{-2/3}$, which still exceeds $1.2$ for a million lions.

\section{An impurity in a one-dimensional trapped Bose gas} \label{S:sec-impurity}

The lamb and the lions have an exact quantum counterpart which is of current interest for ultracold atoms, and which we mentioned in the conclusion of the main text.

\emph{Mapping.} Consider the lamb and $N$ lions, and let $X_i = x_i/\sqrt{t}$ and $\tau = \ln t$. Since $\dd X_i = - \frac{1}{2} X_i \, \dd \tau + \sqrt{2D} \, \dd W_i$, where the $W_i(\tau)$ are independent Wiener processes, the density $P(\vb{X}, \tau)$ of the configurations in which the lamb is still alive satisfies
\begin{equation} \label{S:OU}
    \pdv{P}{\tau} = D \laplacian P + \frac{1}{2} \nabla \cdot \left( \vb{X} P \right) \;,
\end{equation}
and vanishes on the walls $X_0 = X_i$. The substitution $P = e^{-\vb{X}^2/8D} \psi$ turns it into a Schr\"odinger equation in imaginary time, $- \partial_\tau \psi = (H - E_{\rm free}) \psi$, with
\begin{equation} \label{S:trap-H}
    H = - D \laplacian + \frac{\vb{X}^2}{16 D} \;, \mbox{~~and~~} E_{\rm free} = \frac{N + 1}{4} \;.
\end{equation}
With $\hbar = 1$, $H$ is the Hamiltonian of $N + 1$ particles of mass $M = 1/2D$ in a harmonic trap of frequency $\omega_0 = \frac{1}{2}$, and $E_{\rm free} = (N+1) \omega_0/2$ is its ground-state energy in the absence of walls. The condition $\psi = 0$ on the walls is the limit $g \to \infty$ of a contact repulsion $g \sum_{i \geq 1} \delta(X_0 - X_i)$ between the lamb and each lion, while the lions do not interact with one another. When all the pairs repel infinitely, and for the mass ratios such that the walls are the mirrors of a reflection group, the spectrum is exactly solvable \cite{HODVJZ17}. Here only the mirrors which involve the impurity are walls. Since a wall cannot be crossed, the number of lions on each side of the lamb is conserved, and the lamb and lions problem is the sector in which all the lions are on the same side. The survival probability decays as $e^{-(E_0 - E_{\rm free}) \tau} = t^{-(E_0 - E_{\rm free})}$, where $E_0$ is the lowest level of $H$ in this sector. Restoring the units we get
\begin{equation} \label{S:beta-E}
    \beta_N = \frac{E_0 - E_{\rm free}}{2 \hbar \omega_0} \;.
\end{equation}
This is consistent with Sec.~\ref{S:sec-cone}. Once the center of mass is removed, $H$ is an isotropic harmonic oscillator in $N$ dimensions restricted to the cone, with the levels $\hbar \omega_0 (2n + a + N/2)$ for $n = 0, 1, \cdots$, where $a (a + N - 2)$ runs over the eigenvalues of the Laplacian on the part of the unit sphere cut by the cone. The lowest one, with $a = 2 \beta_N$, is to be compared with $\hbar \omega_0 N/2$ without walls. For three lions $a = \nu_1$: the relative motion has the noninteger angular momentum $\nu_1$.

The ground state is symmetric in the lions, which can therefore be regarded as identical bosons. In the language of ultracold atoms the lamb is an impurity which repels, with an infinite contact interaction, $N$ noninteracting bosons of the same mass in the same harmonic trap, and
\begin{equation} \label{S:impurity-energy}
    \mathcal{E}_N \equiv E_N(g = \infty) - E_N(g = 0) = 2 \beta_N \, \hbar \omega_0
\end{equation}
is the energy of the impurity, $E_N(g)$ being the ground-state energy of the $N + 1$ atoms. The two sectors in which the impurity is on the right or on the left of all the bosons are mirror images of each other, which is the two-fold degeneracy of the ground state at infinite repulsion noted in Ref.~\cite{DVZ15}. The sectors with bosons on both sides of the impurity lie higher: for three bosons, the last line of Table~\ref{tab:orderings} gives $\mathcal{E} = 4.14321 \, \hbar \omega_0$ when one boson is on one side and two are on the other.

\emph{Checks.} For a single boson $\mathcal{E}_1 = \hbar \omega_0$: two impenetrable particles have the energy of two free fermions. If all the walkers were mutually vicious, the same mapping would give $n$ free fermions with $E_0 - E_{\rm free} = \hbar \omega_0 \, n (n-1)/2$, i.e. the known exponent $n (n-1)/4$ of $n$ vicious walkers \cite{Fisher84,HF84}. For two bosons $\mathcal{E}_2 = \frac{3}{2} \hbar \omega_0$, which is the exact result of Ref.~\cite{ZVFJV14}. Table~\ref{S:tab-impurity} compares Eq.~(\ref{S:impurity-energy}) with the exact diagonalizations of Ref.~\cite{RIJ24}, where this quantity is denoted $\mu$. They were performed at a large but finite coupling ($g = 1000$ in oscillator units) and in a truncated basis, with 90, 45, 30, 20, 15 and 15 oscillator levels per particle for $N = 2, \cdots, 7$ bosons. The finite coupling lowers the energy by $K/g$, with $K$ between 1.1 and 1.9 \cite{RIJ24}, i.e. by less than 0.002, and the truncation raises it. The two sets of values agree to within 0.002 up to five bosons, and to within $0.6\%$ for six and seven bosons, for which the basis is the smallest. This is an independent check of the exponents of Table~\ref{S:tab-ladder}. For three bosons the solution of the main text gives the exact value $\mathcal{E}_3 = \nu_1 \hbar \omega_0$.

\begin{table}[h]
\caption{Energy of an impurity with infinite contact repulsion in a harmonically trapped gas of $N$ noninteracting bosons, in units of $\hbar \omega_0$. First line: Eq.~(\ref{S:impurity-energy}) with the exponents of the main text and of Table~\ref{S:tab-ladder}. Second line: exact diagonalization at $g = 1000$ (Table~1 of Ref.~\cite{RIJ24}).}
\label{S:tab-impurity}
\begin{ruledtabular}
\begin{tabular}{ccccccc}
$N$ & 2 & 3 & 4 & 5 & 6 & 7\\
\hline
$2 \beta_N$ & $3/2$ & 1.8257570 & 2.0645357 & 2.2517315 & 2.405014 & 2.53440\\
Ref.~\cite{RIJ24} & 1.499 & 1.826 & 2.064 & 2.253 & 2.415 & 2.549\\
\end{tabular}
\end{ruledtabular}
\end{table}

\emph{Many bosons.} The law $\beta_N \simeq \frac{1}{4} \ln N$ \cite{KR96,LS02} translates into
\begin{equation} \label{S:impurity-large-N}
    \mathcal{E}_N \simeq \frac{1}{2} \hbar \omega_0 \ln N \mbox{~~when~~} N \to \infty \;,
\end{equation}
and has a simple meaning in this language. The bosons occupy the ground state of the trap, whose density is a Gaussian of variance $\hbar/2 M \omega_0$, so the outermost of $N$ bosons is typically found at $x_{\max}^2 \simeq (\hbar/M \omega_0) \ln N$. The impurity is expelled beyond this point, where its potential energy is $\frac{1}{2} M \omega_0^2 x_{\max}^2 = \frac{1}{2} \hbar \omega_0 \ln N$. This is the quantum form of the argument of Ref.~\cite{KR96} according to which the lamb only has to stay ahead of the leading lion. The energy of the impurity therefore grows without bound, albeit slowly, with the size of the gas. In Ref.~\cite{RIJ24} the values for $N \leq 7$ were fitted to a form which saturates at $6.7 \, \hbar \omega_0$ for large $N$. Over this range the two forms are indeed hard to distinguish, but the limit $\beta_N/\ln N \to \frac{1}{4}$ proved in Ref.~\cite{LS02} excludes a saturation.